\documentclass[twocolumn,showpacs,amsmath,superscriptaddress]{revtex4}
\usepackage{graphicx}

\newcommand{\rmd}{{\rm d}}
\newcommand{\rme}{{\rm e}}
\newcommand{\rmi}{{\rm i}}

\begin{document}

\title{Effective mass in cavity QED}

\author{Jonas Larson}
\email{jolarson@kth.se}
\affiliation{
Laser Physics and Quantum Optics, Albanova, Royal Institute of Technology (KTH),
SE-10691 Stockholm, Sweden}
\author{Janne Salo}
\affiliation{
Laser Physics and Quantum Optics, Albanova, Royal Institute of Technology (KTH),
SE-10691 Stockholm, Sweden}
\affiliation{Helsinki University of Technology, Materials Physics
Laboratory,
02015 HUT, Finland}
\author{Stig Stenholm}
\affiliation{
Laser Physics and Quantum Optics, Albanova, Royal Institute of Technology (KTH),
SE-10691 Stockholm, Sweden}

\date{\today}

\begin{abstract}
We consider propagation of a two-level atom coupled to one electro-magnetic mode of a 
high-Q cavity. The atomic center-of-mass motion is treated quantum mechanically and 
we use a standing wave shape for the mode. The periodicity of the Hamiltonian leads to 
a spectrum consisting of bands and gaps, which is studied from a Floquet point of view. 
Based on the band theory we introduce a set of effective mass parameters that 
approximately describe the effect of the cavity on the atomic motion, with the emphasis 
on one associated with the group velocity and on another one that coincides with the 
conventional effective mass. Propagation of initially Gaussian wave packets is also 
studied using numerical simulations and the mass parameters extracted thereof are 
compared with those predicted by the Floquet theory. Scattering and transmission of the 
wave packet against the cavity are further analyzed, and the constraints for the effective 
mass approach to be valid are discussed in detail.
\end{abstract}


\maketitle

\section{Introduction}
\label{intro} Cavity quantum electro dynamics (QED)~\cite{qed} has
experienced a tremendous progress during the last decades. In
experiments where atoms interact with a cavity field, the
lifetimes of both the cavity and atomic states can be made rather
long, up to tens of milliseconds. This makes it possible to carry
out several operations on the combined system before decoherence
plays an influential role. It is also possible to single out a unique
atomic transition to interact with only one cavity mode, implying
that only two atomic states $|\uparrow\rangle$ and $|\downarrow\rangle$
together with one electro-magnetic mode $|n\rangle$ need to be taken into account,
while other atomic states and modes are neglected. In such
situations, the Jaynes-Cummings (JC) model~\cite{jc1,jc2} has
proven to be remarkably well suited for describing the coherent
interaction. Cavity QED has thus become one of the candidates for
implementing quantum information processing, see for example
\cite{qc1,qc2,qc3,qc4} and it has also turned out to be a very
useful tool for studying purely quantum mechanical phenomena, such
as sub-poissonian Fock states~\cite{fock} and
Schr\"odinger cat states~\cite{cat}.

The simple JC model is not, however, always valid. For example, if
the atom's kinetic energy is of the same order of magnitude as the
atom-cavity interaction energy, the dynamics is significantly
changed~\cite{mazer0,mazer1}. Thus, for very cold atoms, the
kinetic energy term for the atomic center-of-mass motion must be
treated quantum mechanically. In the standard JC model, the atom
is either assumed to stay still relative to the cavity mode, or to
have a large kinetic energy; in both cases the atomic motion is
described classically and the kinetic-energy term may be excluded from
the Hamiltonian. Another simplification of the JC model is that
the spatial shape of the cavity mode is not taken into account and it is
assumed constant. This is, of course, not always valid, since an
atom traversing a cavity will see a mode that varies with respect
to the atomic position. The mode variation is given by the
particular shape of the electric field and is, therefore, space dependent.
The proper approach in such a case is to introduce an atom-field coupling that
is position-dependent $g(x)=\bar{d}\cdot\bar{E}(x)/\hbar$, where $\bar{d}$
is the dipole moment of the atomic transition and $\bar{E}(x)$ is the
electric field  of the cavity mode involved. For a
smooth coupling and small velocities, the atoms see the cavity
mode as an effective potential which is, in the adiabatic limit,
given by $\pm\sqrt{\Delta^2/4+g^2(x)}$, here $\Delta$ is the
atom-cavity detuning. Consequently, the atom experiences an effective force
from the potential and it may, for instance, be reflected or
transmitted by the cavity~\cite{refl1,refl2,refl3} or even trapped
inside it~\cite{trap1,trap2,trap3,trap4,trap4b,trap5,trap6}. The
situation in which the atom experiences both an effective cavity
potential and an external potential has also been discussed
\cite{ext}. Today it is possible to trap ions inside cavities even
using external traps~\cite{atcavtrap}, which open up new
possibilities for realizing certain desirable systems.

The shape of the cavity mode depends on the boundaries of the
cavity, the most commonly considered being Gaussian, standing
wave, travelling waves in ring cavities, and whispering-gallery
modes. We consider here a standing wave mode with a wave number
$q$, $g(x)=g_0\cos(qx)$, where $g_0$ is the scaled strength of the
coupling. For such a system, an extended JC model including atomic
centre of mass motion and a standing wave coupling has been
studied in a large number of papers, here we just mention a few.
The dynamics has been analysed in, for example,
\cite{dyn1,dyn2,dyn3,dyn4,dyn5,dyn6,appdyn1,appdyn2,appdyn3,appdyn4},
while in~\cite{appdyn1,appdyn2,appdyn3,appdyn4}, approximation
methods are used, such as Raman-Nath, Bragg, tight-binding or
large detuning. In the Raman-Nath approximation, the kinetic
energy term is neglected, and this has been assumed in
Refs.~\cite{rm1,rm2,rm3}, where effects from various measurements
on the field or the atom have been studied.

Clearly, with a standing wave coupling, the Hamiltonian is
periodic with period $\lambda=2\pi/q$. The spectra of periodic
Hamiltonians are known to consist of allowed energies in forms of
bands, separated by forbidden gaps. They are most commonly treated
using the Floquet theory, which has been done, for example, in
Refs.~\cite{dyn1,dyn4,appdyn2}. An interesting observation is that
the Brilluin zone is now twice as wide as in the usual case for
one dimensional periodic Hamiltonians. This derives from the fact
that the two atomic levels are coupled to the motion, in contrast to electrons in solid
crystals where electronic spin flips are not coupled to the
lattice potential. In an atom-cavity system, every time the atom
absorbs or emits a photon and gets a momentum 'kick', its internal
state $|\pm\rangle$ is also changed. Hence, in the rotating wave
approximation, an emission (absorption) must take place between
two absorptions (emissions). The symmetry of the
system is, therefore, generated by displacement of half the spatial
period accompanied by an atomic inversion (performing this twice
yields the spatial periodicity of the Hamiltonian), which renders
the Brillouin zone (in momentum space) twice as wide.

In solid state physics, it has proved useful to describe an
electron propagating within a periodic structure in terms of a
dispersion relation $E=E^\nu(k)$ where $k$ is called the
quasi-momentum of the state and $\nu$ is an index for the
electronic band; in this picture the electron is considered to
move freely, with its propagation characteristics defined by the
dispersion relations. Here, as below, we set Planck's constant $\hbar=1$. If the electronic wave function is
represented by a (Gaussian) wave packet centred around (quasi-)
momentum $k_0$, its propagation velocity, i.e., group velocity, is
given by $v_{{\rm g}}=\partial E^\nu(k)/\partial k|_{k=k_0}$. Thus, the
velocity of the electron is no longer determined by its original
mass, such that $v=k/m$, but by a renormalized mass defined as
$m_1=k/v_{{\rm g}}$, which depends on the dominant quasi-momentum of the
state. As the tangent of the dispersion curve gives the group
velocity, which is related to the free space velocity through the
ration $m/m_1$, the curvature determines the amount of spreading
of the Gaussian wave packet and likewise it defines another mass
parameter $m_2=\left(\frac{\partial^2E}{\partial
k^2}\right)^{-1}$. In this paper we study the dynamics of a
two-level atom interacting with a standing wave cavity mode and
discuss the effect of masses $m_1$ and $m_2$, replacing the
original free mass of the atom.

In ordinary QED, the assignment of mass to electrons is an
essential part of the renormalization program, where the mass may
be considered shifted by the presence of the zero-point energy of
the vacuum; the fact that formally infinite entities are
manipulated does not invalidate the general picture. Likewise, one
may expect that the presence of a finite energy in the field may
give its own contribution to the renormalized quantities, in
particular the mass. This quantity is usually considered to be too
small to have any observable consequences. In a cavity, on the
other hand, the coupling of an atom to the cavity modes is
enhanced, and it may be possible to interrogate the effect of the
field on the mass.

In most setups for atom-cavity QED experiments, the atom is
prepared in some initial state outside the cavity and is then
allowed to propagate through the cavity field. Provided that the
photonic wavelength $\lambda$ is small compared with the cavity
length, the system can be treated approximately as periodic, and
the results of the Floquet theory are applicable. If, for
instance, the atom is prepared with a kinetic energy that lies in
a forbidden energy gap, it cannot enter the cavity but must be
reflected from it, possibly with a flipped internal atom-field
state $|\pm\rangle$, as will be shown below. When the kinetic
energy falls within the allowed energy bands, the atom will
traverse the cavity with a (group) velocity $v_{{\rm g}}=k/m_1$. We also
simulate wave packet propagation in these situations using the
split operator method. The results obtained from the Floquet
theory and the wave packet propagations are compared. Since the
mass parameters $m_1$ and $m_2$ depend on the effective coupling
$g_0\sqrt{n}$, where $n$ is the photon number, a measurement of
$m_1$ or $m_2$ also yields indirectly the photon number inside the
cavity. The reflection and transmission of atoms against the
cavity may also be used for state preparation or 'Stern-Gerlach'
type of measurements between different internal orthogonal states.

The paper is outlined as follows: First, in Section~\ref{s2}, the
Hamiltonian describing the dynamics is introduced and solved
numerically for the eigenenergies and eigenstates in accordance
with the Floquet theory. The bare and dressed states of the system
are presented and analytic approximations for the lowest band is
obtained. The more illustrative approach of wave packet simulations
is considered in \ref{s3} and the effective masses $m_1$ and $m_2$
are defined. Both the propagation of Gaussian bare and Gaussian
dressed states are discussed and, in order to get a deeper
understanding of these two cases, we analyze the comparison
between bare and dressed states. The masses $m_1$ and $m_2$ are
extracted numerically and compared between bare and dressed states
wave packet propagation and also with the masses obtained from the
Floquet theory. Further it is shown with simulations how atoms
may be reflected or transmitted by the cavity mode and we discuss
possible applications for state preparation and state
measurements. Finally, in \ref{s4} we conclude with a discussion
of the results and possibilities to observe the mass in realistic
experiments.


\section{The Floquet approach}
\label{s2}

We describe the atom-cavity system with a Jaynes-Cummings
model~\cite{jc1} that takes into account two atomic levels,
coupled to a single field mode in the rotating wave and dipole
approximations; two essential parameters are the atom-field
coupling $g_0$ and the detuning $\Delta$ between the atomic
transition frequency $\omega$ and mode frequency $\Omega$.
Moreover, the field mode is assumed to be a standing wave along
the cavity axis $x$, and the parallel atomic motion is
quantized. The spectrum of the Hamiltonian is obtained using the
Floquet theory, and it has a band structure with Brillouin zones
twice as wide as for one-level particals~\cite{dyn1,dyn4,appdyn2}.
The effects of the band and band-gaps will be discussed in
Section~\ref{s3}, where we consider physically realistic situations.

\subsection{Jaynes-Cummings model for a moving atom}\label{ss2a}

The Jaynes-Cummings model~\cite{jc1} describes the interaction
between a two-level atom and a single field mode. As mentioned in
the Introduction above, the atomic center-of-mass motion has been
ignored and the parameters are assumed to be independent of the
atomic position in the original Jaynes-Cummings model; these
conditions are not, however, always satisfied in realistic
atom-cavity experiment. As the atom traverses the cavity, the
shape of the coupling will be governed by the cavity mode
structure. For a standing wave mode, with wave number
$q=2\pi/\lambda$, the mode is a given by $g(x)=g_0\cos(qx)$. In
most studies and experiments, the atomic velocity is large enough
that it can be described classically; thus its energy merely
adds a $c$-number to the Hamiltonian. In such situations, assuming
the atom to be point-like, the position operator $x$ can be
replaced by the classical center-of-mass position moving with the
velocity $v$, i.e., $x=vt$, see~\cite{jonas2,schlicher}.
For cold atoms, however, the center
of mass motion must be considered quantum mechanically
\cite{mazer0,mazer1} and the kinetic energy operator term must be
included in the original Hamiltonian. In many of the studies where
the kinetic energy term is included, however, the system is
simplified by adiabatic elimination of the excited state in the
limit of large detuning~\cite{stigadd1,stigadd2}.

The extended Jaynes-Cummings Hamiltonian, with standing wave mode
structure and quantized atomic motion, becomes
\begin{equation}
H=\frac{P^2}{2m}+\frac{1}{2}\hbar\tilde{\Delta}\sigma_3+\hbar \tilde{g}(X)\left( a\sigma^++a^{\dagger}\sigma^-\right);
\end{equation}
Here the tilde notation ($\sim$) indicates the original dimensional
variables. Also $m$ is the atomic 'free' mass, capital $P$ and $X$ are the
momentum and position operators, $a$ and $a^{\dagger}$ boson
annihilation and creation operators for the cavity mode, and the
$\sigma$-operators are the Pauli matrices acting on the internal
two states of the atom. Note that we only consider quantized
motion in one dimension along the cavity axis. The coupling will
be taken as $\tilde{g}(X)=2\tilde{g_0}\cos(\tilde{q}X)$. The
evolution Hamiltonian is written in the interaction picture with
respect to the 'free Hamiltonian'
$H_0=\hbar\Omega(\frac{1}{2}\sigma_3+a^\dagger a+\frac{1}{2})$ and
$\tilde\Delta=\omega-\Omega$ is the atom-cavity detuning.

Since the Hamiltonian has been given in the rotating-wave
approximation, the total number of excitations in the system
$N=a^{\dagger}a+\frac{1}{2}\sigma_3$ is a conserved quantity and
the dynamics therefore splits up into separated decoupled
subsystems for each number of excitations. In the joint Hilbert
space of the internal atomic state and the cavity mode, we define
the basis states as
\begin{equation}
\begin{array}{cc}
|+\rangle &= \left[\begin{array}{c}1\\0\end{array}\right]=|\uparrow,n-1\rangle\\ \\
|-\rangle &= \left[\begin{array}{c}0\\1\end{array}\right]=|\downarrow,n\rangle,
\end{array}
\end{equation}
where the atomic states $|\uparrow\rangle$ and
$|\downarrow\rangle$ are the atomic upper state and lower states
of the transition, and $|n\rangle$ are the cavity mode Fock
states. Using this basis and scaled parameters, the Hamiltonian
assumes the form
\begin{equation}\label{ham1}
H=-\frac{1}{2}\frac{\partial^2}{\partial x^2}+\left[\begin{array}{cc} \frac{\Delta}{2} & \sqrt{n} g(x) \\ \sqrt{n} g(x) & -\frac{\Delta}{2}\end{array}\right],
\end{equation}
where, after introducing a characteristic time and length scale
$T_s$ and $X_s$, the scaled parameters are expressed in terms of
the old ones according to
\begin{equation}
\begin{array}{ccc}
g=T_s\tilde{g}, & & \Delta=T_s\tilde{\Delta}=T_s(\omega-\Omega), \\ \\ x=\frac{X}{X_s}, & q=\tilde{q}X_s, & T_s=\frac{mX_s^2}{\hbar},
\end{array}
\end{equation}
and $n$ is the photon number. We will take the photon momentum $\tilde{q}$
to define the characteristic length scale as $X_s=1/\tilde{q}$. We shall consistently
indicate $q$ in all equations below, but use the numerical value $q=1$ in all the
figures in accordance with the chosen length scale. In this
way, momenta $k$ will be given in units of q and the relevant
parameters of the model are $g_0/q^2$ and $\Delta/q^2$. 
In most of the following analysis, we will assume the atom
to be initially in its ground state $|\downarrow\rangle$ and the
mode to contain one single photon. We should emphasize that the
Hamiltonian (\ref{ham1}) then becomes identical to the on
describing a two-level atom interacting with a classical standing
wave field. Therefor, our model may not only be used for
describing atom-cavity QED dynamics, but also the interaction
between two-level atoms and classical fields, for example, if the
cavity is driven by a classical source or the field is given by a
laser beam. More general situations of the quantized field could
be considered in a straightforward generalization, but in this
paper we only discuss the basic features and keep the model as
simple as possible.

An interesting observation is that for zero detuning $\Delta=0$, the unitary operator
\begin{equation}
U=\frac{1}{\sqrt{2}}\left[\begin{array}{cc} 1 & 1 \\ 1 & -1\end{array}\right]
\end{equation}
decouples the system into two ordinary one-dimensional
Schr\"odinger equations with potentials $V_{\pm}(x)=\pm2\cos(qx)$;
these equations are known as Mathieu equations~\cite{math}.

Due to the spatial periodicity $\lambda$ of the cavity mode, the operator
\begin{equation}
T={\rm e}^{i\lambda p},
\end{equation}
with $\lambda = 2\pi/q$, commutes with the Hamiltonian
(\ref{ham1}); this symmetry property is the background for the
Floquet theory. Another, slightly less obvious symmetry is
associated with the operator
\begin{equation}
I=\sigma_3{\rm e}^{i\frac{\lambda}{2}p}
\end{equation}
that also commutes with the Hamiltonian~\cite{dyn1}. This is a
'half-period' displacement combined with an atomic inversion
and it includes the first symmetry operation since
$T=I^2$. Consequently, the first Brillouin zone is within
$-q<k<q$, and not within $\pm q/2$, as implied by $T$ alone.
Physically this derives from the fact that every absorption or
emission of a photon flips the internal state
$|\pm\rangle\rightarrow|\mp\rangle$, while after absorption +
emission (or vice versa) the internal atomic state remains
unchanged. The center-of-mass momentum in the two step process
will either be the same or shifted by $\pm 2q$, depending on the
direction of the emitted/absorbed photons.Note that the first Brillouin zone has occasionally been defined to extend within $-\frac{q}{2}<k\le \frac{q}{2}$, which produces two sets of dispersion curves, one for each internal state $|\pm\rangle$, see Refs.~\cite{dyn1,kolovsky}.

\subsection{Energy band structure}

Due to the conservation of the total excitation number
$N=a^\dagger a + \frac{1}{2}\sigma_3$, the system Hilbert space $\mathcal{H}$
may be represented as a direct sum of $\mathcal{H}_1$ and $\mathcal{H}_2$
that are not coupled by the Hamiltonia. We limit the discussion to $\mathcal{H}_1$
that is spanned by the {\it bare states} 
\begin{equation}\label{bare}
|\psi_{\mu}(k)\rangle=\left\{\begin{array}{lll}|k+\mu q\rangle|-\rangle & & \mathrm{\mu\,\,\,even} \\ |k+\mu q\rangle|+\rangle & & \mathrm{\mu\,\,\,odd}\end{array}\right.
\end{equation}

which are energy eigenstates of the system in the absence of interaction, with their energies given by $\mathcal{E}^\mu=\frac{1}{2m}(k+\mu q)^2- (-1)^\mu\frac{\Delta}{2}$. The quasimomentum $k$ is here limited into the first Brillouin zone $-q< k \le q$ and the integer index $\mu$ denotes the Brillouin zone or, equivalently, the energy band. Thus the physical momenta of the bare states have well-defined values $k+\mu q$ and, in particular, the internal $|-\rangle$ state with zero-momentum is given by $|\psi_{0}(0)\rangle$ while the $|+\rangle$ state, by $|\psi_{-1}(q)\rangle$.

The energy eigenstates of the interacting Hamiltonian given by Eq.~(\ref{ham1}),
\begin{equation}
\begin{array}{cccc}
H|\phi_{\nu}(k)\rangle=E^{\nu}(k)|\phi_{\nu}(k)\rangle, & & & \nu=1,2,3,...
\end{array},
\end{equation}
are called {\em dressed states} and they can be expressed as linear combinations of the bare states
\begin{equation}\label{dress1}
|\phi_{\nu}(k)\rangle=\sum_{\mu=-\infty}^{\infty}\,c_{\mu}^{\nu}(k)\,|\psi_{\mu}(k)\rangle.
\end{equation}
Each dressed state is assigned to some energy band (Brillouin
zone) $\nu$, which is numbered $1,2,3,...$ for increasing energy, and is also indexed with a continuous variable $k$,
which is now called the quasi-momentum; the entire quantum states
contain all momenta $k+\mu q$. The functional dependence of the
energy eigenvalue on the quasi-momentum $E^{\nu}(k)$ is called the
dispersion curve, assigned to each Brillouin zone. The dressed
states for each quasi-momentum are obtained by solving the secular
equation given by the infinite matrix Hamiltonian
\begin{equation}\label{matrix}
\tiny
\left[
\begin{array}{ccccccc}
\ddots & \vdots & \vdots & \vdots & \vdots & \vdots &  \\
\ldots & \frac{(k-2q)^2}{2}\!-\!\frac{\Delta}{2}  & g_0 & 0 & 0 & 0 & \ldots \\
\ldots & g_0 &  \frac{(k-q)^2}{2}\!+\!\frac{\Delta}{2} & g_0 & 0 & 0 & \ldots \\
\ldots & 0 & g_0 &  \frac{k^2}{2}\!-\!\frac{\Delta}{2} & g_0 & 0 & \ldots \\
\ldots & 0 & 0 & g_0 &  \frac{(k+q)^2}{2}\!+\!\frac{\Delta}{2} & g_0 & \ldots \\
\ldots & 0 & 0 & 0 & g_0 &  \frac{(k+2q)^2}{2}\!-\!\frac{\Delta}{2} & \ldots \\
 & \vdots & \vdots & \vdots & \vdots & \vdots & \ddots
\end{array}
\right].
\end{equation}
The eigenvalues and eigenvectors of this infinite Hamiltonian are not
known in the general case but approximate analytical results may
be found, see Refs.~\cite{appdyn1,appdyn2,appdyn3,appdyn4}. For
example, one interesting approximation, related to the Raman-Nath
limit, is when the $q^2$-terms are neglected and it may be solved
analytically. Here we will not discuss these approximations, but
first solve the problem numerically and then make different
perturbative expansions for the eigenvalues.

\begin{figure}[ht]
\centerline{\includegraphics[width=8cm]{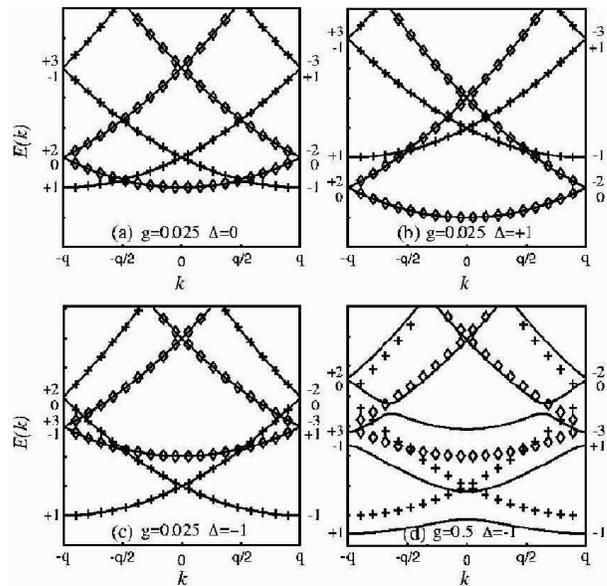}}
\caption[bandstructure]{\label{fig1} The lowest lying bands of the
Hamiltonian given by Eq.~(\ref{ham1}) for the first Brillouin
zone. The dressed energy bands are marked on the $y$-axis with the dominant bare state index
$\mu$. Crosses shows the bare energy bands for excited states $|+\rangle$ and diamonds bands for ground bare states $|-\rangle$. In the last figure (d), the coupling is so strong that the bare and dressed energies starts differ considerable. The parameters are given on top of each figure.}
\end{figure}

In order to solve the problem numerically, the Hamiltonian has to
be truncated at some dimension $n$. For small $k$ and relatively low bands, this $n$ should be chosen odd,
in order to be consistent with coupling to equal number of states
in both 'directions' from a given initial state. For $n=1$,  we
obtain the bare eigenenergy, for $n=3$, the bare state
$|\psi_0^k\rangle$ couples to the states $|\psi_{\pm1}^k\rangle$
and so on. In Figs.~\ref{fig1} (a)-(d) we show the
lowest-lying bands for the first Brillouin zone, obtained
numerically with $n=201$, for the parameters (a) $\Delta=0$,
(b) $\Delta=1$ and (c) and (d) $\Delta=-1$. In (d) the coupling is 20 times as large, making the bare and dressed energies differ significantly. On the $y$-axis
the dressed bands are labelled with the corresponding dominant $\mu$-value that the
bare eigenenergies would have had in the limit of weak coupling, and diamonds indicate a bare
lower state $|-\rangle$ energies and crosses bare states $|+\rangle$ energies. When
adding the coupling, the crossings become 'avoided'. Note how the
gap size decreases with the band index $\nu$, indicating that the
state is more weakly coupled to far-away lying states. The
crossings between even-even $\mu$ or odd-odd $\mu$ are called
Bragg resonances and between odd-even or even-odd $\mu$ Doppleron
resonances~\cite{dyn4}.

\begin{figure}[ht]
\centerline{\includegraphics[width=7cm]{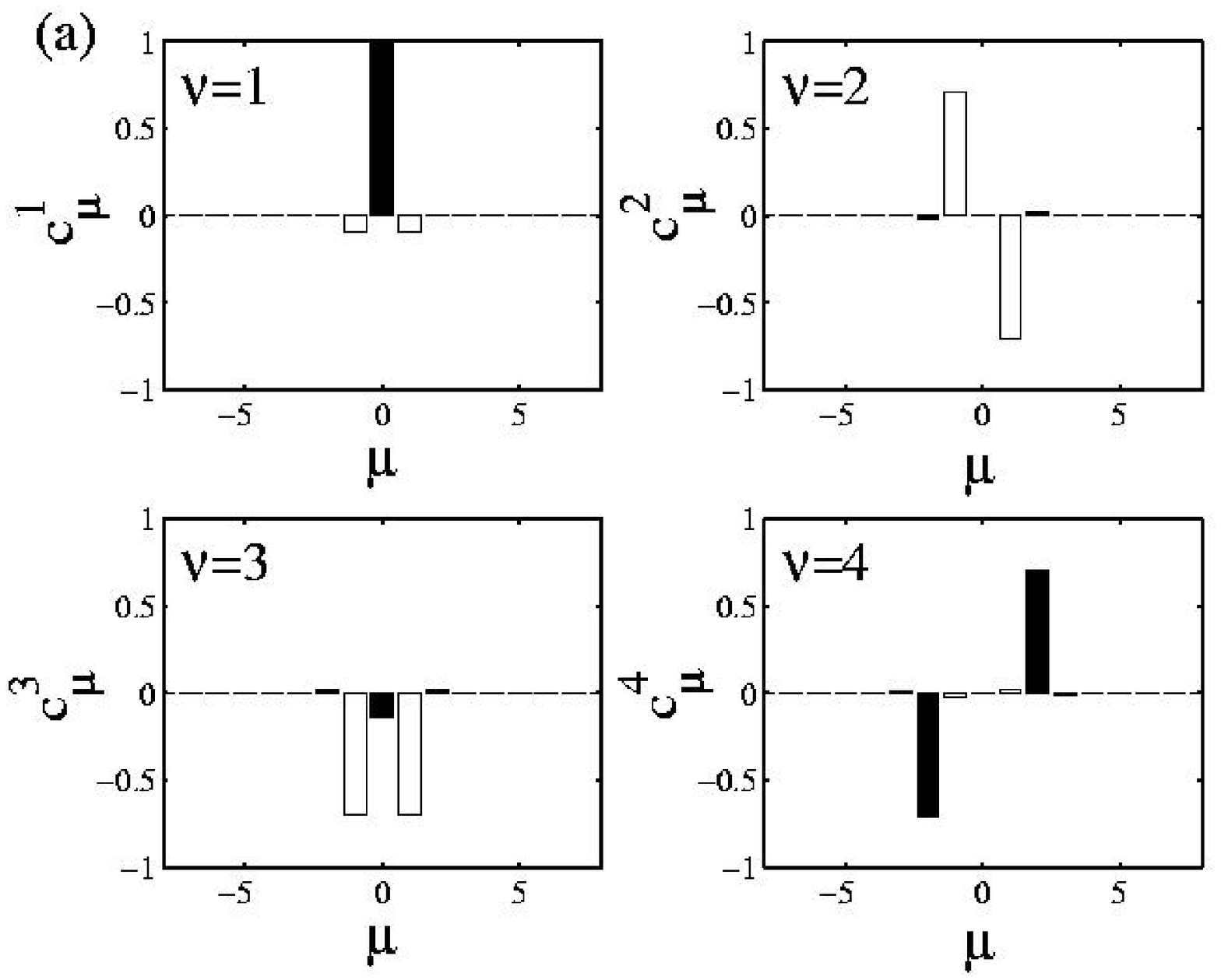}} \vspace{0.5cm}
\centerline{\includegraphics[width=7cm]{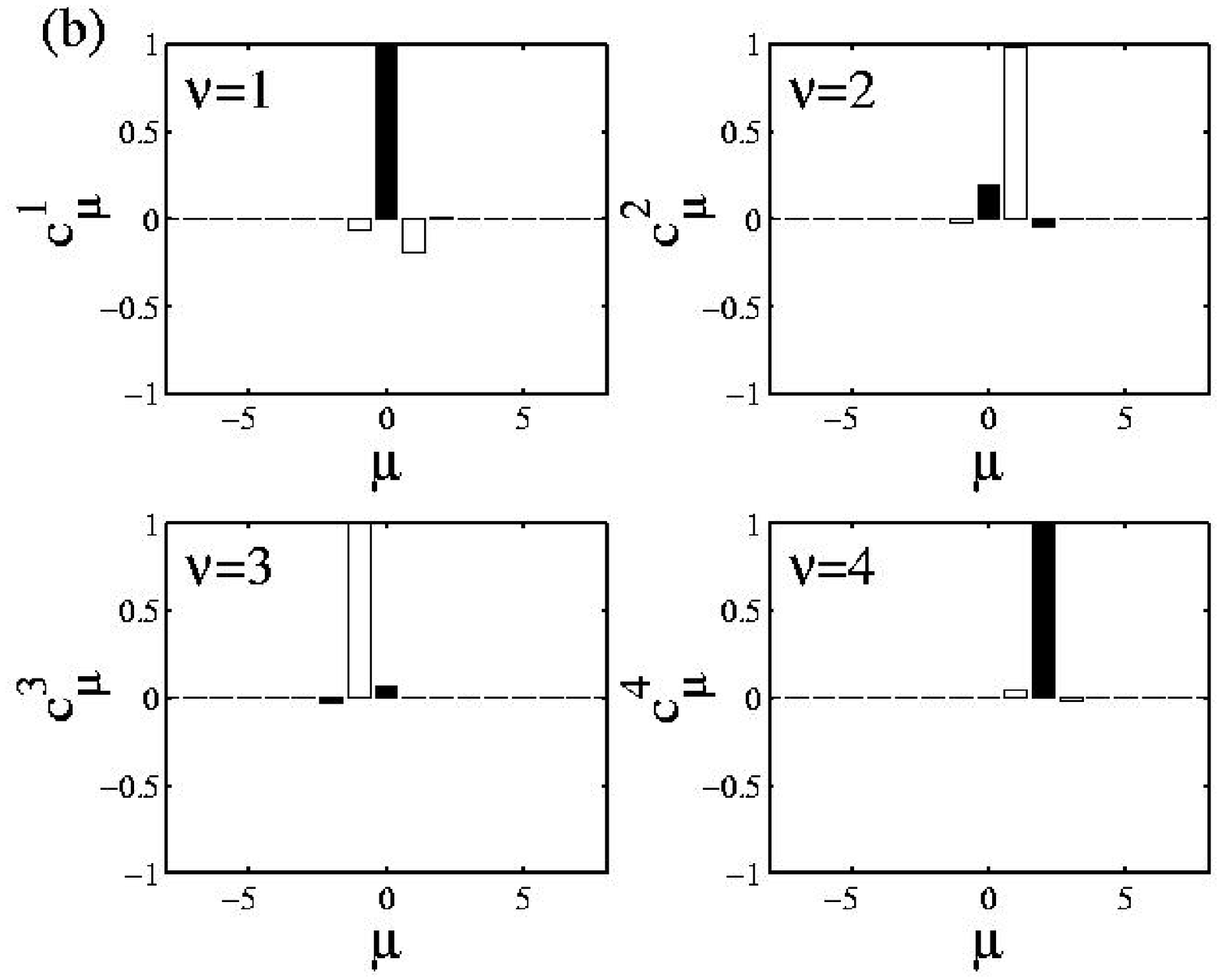}}
\caption[coefficients c_{\mu}^{\nu}]{\label{fig2} The expansion
coefficients $c_{\mu}^{\nu}$ for the four lowest dressed states
$\nu=1,2,3,4$, Eq.~(\ref{dress1}), of bare states $\mu$ for (a)
$k=0$ and (b) $k=1/4$. Black bars corresponds to even $\mu$'s with
ground state atom, and white bars to odd $\mu$'s with excited
atom. The coefficients are symmetric around $\mu=0$ only for
$k_0=0$. The parameters are in both plots the same as for
Fig.~\ref{fig1}: $\Delta=0$ and $g_0=0.05$.}
\end{figure}

In Fig.~\ref{fig2} we illustrate how the presence of the periodic
interaction couples the momentum eigenstates (bare states) into
dressed states. In Fig.~\ref{fig2}(a), the coefficients for the
first four dressed states ($\nu=1,2,3,4$) are plotted as a
function of $\mu$ for zero quasi-momentum $k=0$. Note that the
solutions are either odd or even in $\mu$ and that only the first
one is not 'degenerate' since all other are centered around a
crossing. Figure~\ref{fig2}(b) shows the same coefficients for a
non-zero quasi-momentum  $k_0=q/4$, and the solutions are no
longer symmetric around $\mu=0$. Note that, if any of the
coefficients $c_\mu^{\nu}$ (for each $\nu$) in Eq.~(\ref{dress1})
has an absolute value close to unity, the presence of the periodic
coupling only modifies the properties of a bare state without too
much coupling to other bare states; this is usually the case away
from the crossings.

\subsection{Extraction of the effective parameters}

We now look for an analytical expression for the energy eigenvalues of
the Hamiltonian~(\ref{matrix}) and for the dispersion curves. Denoting
the bare-state
energies by
\begin{equation}
\mathcal{E}^{\mu}(k)
=\frac{(k+\mu q)^2}{2}-(-1)^\mu \frac{\Delta}{2},
\end{equation}
the energy eigenvalue equation $H|\phi\rangle=E|\phi\rangle$ (for
a predefined $k$) can be written as a recursion equation
\begin{equation}\label{recursion}
\mathcal{E}^{\mu} c_{\mu}+g(c_{\mu+1}+c_{\mu-1})
= E c_{\mu}.
\end{equation}
This equation has, naturally, an infinite number of solutions
corresponding to different bands. Truncation of the recursion
symmetrically around some index $\nu$ and elimination of the expansion
coefficients $c_\mu$ yields a continued fraction-like expression
\begin{equation}\label{confraction2}
\begin{array}{ll}
E
= \mathcal{E}^\nu
+ \frac{g^2}{E-\mathcal{E}^{\nu-1}-\frac{g^2}{E-\mathcal{E}^{\nu-2}
+\frac{g^2}{\ldots}}}
+ \frac{g^2}{E-\mathcal{E}^{\nu+1}-\frac{g^2}{E-\mathcal{E}^{\nu+2}
+\frac{g^2}{\ldots}}},
\end{array}
\end{equation}
which has a form of an iteration equation. We assume that this equation
is most applicable in the region where one bare state dominates the
dressed state and the truncation of the Hamiltonian~(\ref{matrix}) or,
equivalently, of the recursion equation~(\ref{recursion}) is performed
symmetrically around this state. Note that
the truncation of the Hamiltonian is also an effective expansion in the
coupling constant $g_0$ since, for a small coupling, each base state only
couples to a few neighboring bare states while, for a large coupling,
many bare states are required to represent a dressed state. This is
not, however, true in the vicinity of level crossings where two bare state couple to each other even over many
intermediate states.

We illustrate the truncation error with 
\begin{equation}\label{error}
\delta(g_0,\Delta;n)=|E(k)-E_{n}(k)|,
\end{equation}
where $E(k)$ represents the exact dispersion curve and $E_{n}(k)$
the one obtained from a truncated $n \times n$ Hamiltonian. 
We emphasize that the error depends only on the parameters
$g_0$ and $\Delta$ due to the chosen length scale; in physical
units they both contain the photon wave number since
$g_0 \propto \tilde{g}_0/\tilde{q}^2$ and
$\Delta \propto \tilde{\Delta}/\tilde{q}^2$.
As already mentioned, we use $q=1$ throughout the paper.

\begin{figure}[ht]
\centerline{\includegraphics[width=8cm]{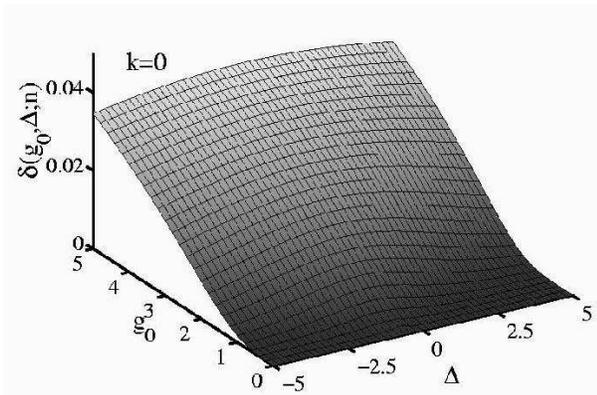}}

\caption[error estimate 1]{\label{fig3} Error estimate
$\delta(g_0,\Delta;n)$ for $n=5$, as defined in Eq.~(\ref{error}).
The quasi-momentum $k=0$. The errors for the $k=1/4$ case is
almost identical.  }
\end{figure}

Figure \ref{fig3} illustrates $\delta(g_0,\Delta;n)$ as function
of $g_0$ and $\Delta$, for $k=0$. An almost identical plot
of the error is obtained for $k=q/4$, therefore it is omitted.
The error increases for large couplings, which is easily understood
since a large coupling means that the initial bare state will
couple more strongly to other bare states and the dimension of the
Hamiltonian must be chosen higher. It is also seen that the
increasing detuning $\Delta$, which makes the diagonal elements in the
Hamiltonian larger, yields smaller errors.

The number of $g_0^2$-terms included in the continued fraction
(\ref{confraction2}) is related to the truncated size of the
Hamiltonian as $(n-1)/2$. Taking any initial value of $E$ and
iterating Eq.~(\ref{confraction2}) it is expected that the iteration
converges to some eigenvalue $E^\nu(k)$ close to the initial value
of $E$. For example, for moderate couplings and away from crossings, starting with $E$ coinciding with a bare energy eigenvalue, the iteration is supposed to converge to the corresponding dressed energy eigenenvalue. Thus, analytical approximate results are obtainable by
truncation the continued fraction to some $(n-1)/2$ terms and
iterate it $j$ times. From numerical investigations of the
validity of these two approximations, we draw the conclusion that
for certain rangers of the parameters, especially for small
couplings $g_0$ and large detunings $\Delta\gg1$, the order of
approximation does not need to be very high away from crossings. It has turned out
that truncating the Hamiltonian to a $5\times5$-matrix and
iterating the continued fraction (\ref{confraction2}) twice gives
an eigenenergy, which is valid for a large range of parameters.
Having an analytical expression for the energy, we can easily 
calculate the mass parameters. Below we will give only the
analytic expression for the case when $k\approx0$, but the same
procedure could be carried out for other cases as well. Since we
have assumed $|k|\ll q$, we expand the eigenenergy around $k=0$,
but we will also expand the result in powers of $g_0^2$. The
result obtained becomes is given by
\begin{equation}
\begin{array}{ccl}\label{appresult2}
E^{(1)}(k) & \approx &
-\frac{\Delta}{2}-\frac{4g_0^2}{q^2+2\Delta}+\frac{4\left(7q^2-
2\Delta\right)g_0^4}{q^2(q^2+2\Delta)}+\mathcal{O}(g_0^6) \\ \\ & &
+\!\left(\!\frac{1}{2}\!-\!\frac{16q^2g_0^2}{(q^2+2\Delta)^3}\!+\!\frac{4(111q^6-
46\Delta q^4-28\Delta^2
q^2-
8\Delta^3)g_0^4}{q^4(q^2+2\Delta)^5}\right. \\ \\ & & +\mathcal{O}(g_0^6)\Big)k^2+\mathcal{O}(k^4).
\end{array}
\end{equation}

\section{Propagation of wave packets}\label{s3}

In this Section, we analyze the propagation of Gaussian dressed
and Gaussian bare states using the effective mass parameters and
compare the results with wave function simulations. We also
discuss the physical difference of initial bare and dressed
states.

Propagation of an initial Gaussian state can be understood in
terms of effective parameters, such as the group velocity and the
effective mass, which depend on the dispersion curve and are
evaluated at the dominant quasi-momentum $k_0$ of the wave packet.
We assume that the wave packet inside the cavity is initially described by a
Gaussian momentum wave function
\begin{equation}
\varphi(k)
= \frac{1}{\sqrt[4]{2\pi\Delta_k^2}}
  {\rm e}^{-\frac{(k-k_0)^2}{4\Delta_k^2}}
\end{equation}
where $\Delta_k$ is the width of the momentum distribution and it
is related to the initial width of the position distribution
according to $\Delta_k=1/(2\pi\Delta_x)$, whence the initial state is a minimum uncertainty state. Within its range, the
energy of the dressed states (belonging to the band $\nu$) can be
expanded as
\begin{equation}
\begin{array}{cl}
E(k)
&\approx E(k_0) + v_{{\rm g}}(k_0)(k-k_0) + \frac{1}{2}\frac{1}{m^*(k_0)}(k-k_0)^2\\ \\
&= E_0
  +\frac{1}{2m_0(k_0)}k_0^2
  +\frac{1}{m_1(k_0)}k_0(k-k_0) \\ \\ &
  +\frac{1}{2m_2(k_0)}(k-k_0)^2,
\end{array}
\label{E_approx}
\end{equation}
with $E_0=E(k=0)$. We have chosen to use mass parameters for each
term in the expansion. Note that, for a free particle, this is
merely an expansion of the energy term $E=\frac{1}{2m}k^2$ around
$k_0$, and all the mass parameters $m_\rmi$ coincide with the
natural mass of the particle.

The three mass parameters in Eq.~(\ref{E_approx}) can now be given
the following interpretations: $m_0$ yields the energy of a
dressed eigenstate as $E(k_0)-E_0=\frac{1}{2m_0}k_0^2$ and, hence,
it defines the phase velocity of the dressed state as
$v_p=E_0/k_0=\frac{1}{2m_0}k_0$, $m_1$ defines the
group-velocity--quasi-momentum relation as $k_0=m_1 v_{{\rm g}}$, and
$m_2$ determines the mass associated with the wave-packet
spreading. Note that, even though we do not consider the case
here, $m_2$ coincides with the conventional effective mass $m^*$
that determines the acceleration caused by an external force
acting on the particle wave packet. The first mass parameter $m_0$
is, however, nonphysical since its value depends on the choice
of zero energy level. It may, however, affect some interference experiment, but we do not consider it here, since it is not expected to effect the propagation.
Equation~(\ref{appresult2}) can be used to deduce
approximate values for $m_1$ and $m_2$ near zero-momentum.

\subsection{Propagation of Gaussian dressed states}

By Gaussian dressed states we mean wave packets centered around
the quasi-momentum $k_0$, where each wave component belongs to the
same energy band of the interacting Hamiltonian,
\begin{equation}
|\Phi_{\nu}(t)\rangle
= \int_{-\infty}^\infty\varphi(k) |\phi_{\nu}(k)\rangle \rme^{-\rmi E^{\nu}(k) t} \rmd k.
\label{dressed_Gaussian}
\end{equation}
In principle, the integral should be limited to the first
Brillouin zone (or to any one Brillouin zone); in this expression it is
assumed that $k_0$ is sufficiently far from its boundaries so that
the momentum distribution $\varphi(k)$ is negligible outside the
zone.

\begin{figure}[ht]
\centerline{\includegraphics[width=8cm]{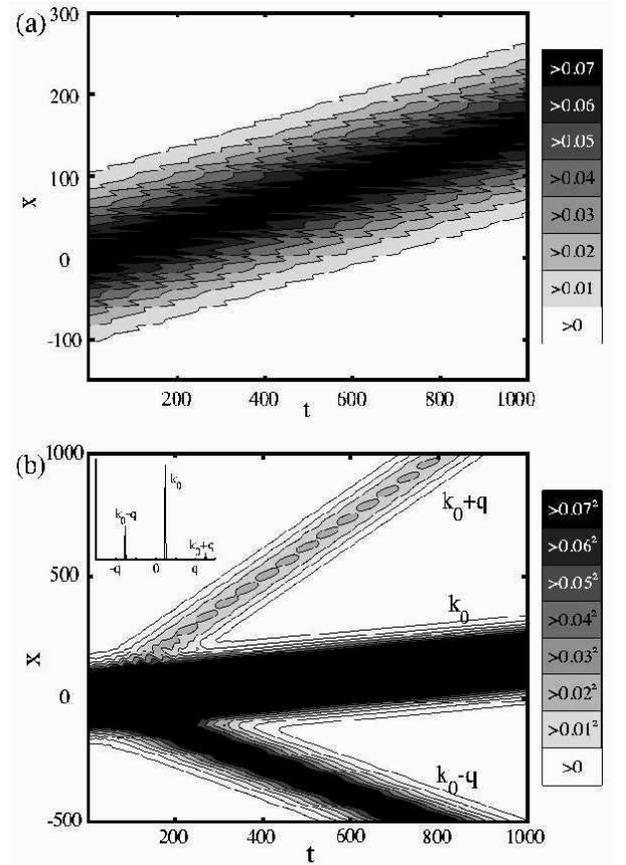}}
\caption[wave packets]{\label{gaussfig} The evolution of (a) an
initial Gaussian dressed state and (b) an initial Gaussian bare
state. The dressed state
stays localized around an average position, while the bare state clearly splits up. The three sub-packets
in (b) correspond to the bare states $|-,k_0\rangle$,
$|+,k_0-q\rangle$ and $|+,k_0+q\rangle$, which is also seen in the
inset showing the final momentum distribution $|\langle
k|\Psi\rangle|^2$. The parameters are the same in both plots,
 $g_0=0.001$, $\Delta=0$, $x_0=0$, $k_0=1/4$ and $\Delta_k^2=1/10000$,
and in (b) the atom is initially in its lower state. The contour-bar shows relative values.}
\end{figure}

With the use of the expansion of the dressed states in terms of
bare states, Eq.~(\ref{dress1}), and of the integral equality

\begin{equation}
\begin{array}{c}
\int_{-\infty}^\infty \varphi(k)\,\rme^{\rmi (kx-E_\nu(k)t)}\,\rmd k  \\ \\ 
\approx
 \frac{1}{\sqrt[4]{2\pi\left( \frac{1}{2\Delta_k}
                             +\frac{\rmi\Delta_kt}{m_2}\right)^2}}
  {\rm e}^{- \frac{{\left( x-v_{{\rm g}} t \right) }^2\,}
          {4\left(\frac{1}{4\Delta_k^2}+\frac{\rmi t}{2m_2}\right) }}
  {\rm e}^{\rmi(k_0 x-E(k_0) t)}
\end{array}
\end{equation}
(the approximate value derives from neglecting higher terms in
Eq.~(\ref{E_approx})), the Gaussian dressed states evolve in time
as
\begin{equation}\label{dresspacket}
\begin{array}{cc}
|\Phi_{\nu}(t)\rangle
&= \frac{1}{\sqrt[4]{2\pi\left( \frac{1}{2\Delta_k}
                             +\frac{\rmi\Delta_kt}{m_2}\right)^2}}
  {\rm e}^{- \frac{{\left( x-v_{{\rm g}} t \right) }^2\,}
    {4\left(\frac{1}{4\Delta_k^2}+\frac{\rmi t}{2m_2}\right) }}
   |\phi_{\nu}(k_0)\rangle
\end{array}
\end{equation}
and has the time-dependent width
$\Delta_x(t)=|\frac{1}{2\Delta_k}+\frac{\rmi \Delta_k t}{m_2}|$.
Here we have further assumed that the expansion coefficients
$c_{\mu}^{\nu}(k)$ remain constant within the Gaussian momentum
distribution. Inclusion of a correction term $c_{\mu}^{\nu}(k)
\approx c_{\mu}^{\nu}(k_0) + d_{\mu}^{\nu}(k_0)(k-k_0)$ gives rise
to an additional term in the integral, but it still has a Gaussian
envelope moving with the same group velocity. In Figs.~\ref{gaussfig} (a) and (b) we show the propagation of initial Gaussian dressed and bare states. The dressed
state stays approximately Gaussian throughout the evolution, while
the initial bare state splits up in three main sub-packets
corresponding to the bare states $|-,k_0\rangle$,
$|+,k_0-q\rangle$ and $|+,k_0+q\rangle$. In order to prepare the
initial dressed states, the coupling amplitude is taken to be time
dependent $g_0(t)$ and is turned on adiabatically from $g_0=0$ up
to the final value $g_0(t)=g_0$. In that way, the state remains
dressed during the turn-on assuming an adiabatic switch on. As
$g_0(t)$ has reached the final value $g_0$, the wave packet has
already broadened, so after the preparation process we do no longer
have $\Delta_x=1/(2\Delta_k)$. The numerical method used for wave
packet simulations will be discussed in Section~\ref{ss3c}.

\subsection{Comparison between bare and dressed states}\label{ss3b}

While in most experimental situations an atom enters a cavity in a
well-defined bare state, the mass parameters defined above characterize a dressed state and are not directly applicable. We
introduce therefore the overlap (often called Fidelity) between bare and dressed
states
\begin{equation}\label{overlap}
F^{\nu,\mu}(k)=|\langle\psi_{\mu}^k|\phi_{\nu}^k\rangle|^2,
\end{equation}
which is the same as the absolute square of the coefficients
$c_\mu^\nu$ in Eq.~(\ref{dress1}) For a relative small coupling and away from crossings, it is clear
which bare state $|\psi_{\mu}^k\rangle$ has the largest overlap
with a given dressed state $|\phi_{\nu}^k\rangle$. However, near
crossings two bare states will become important, but still it is
easy to decide which ones. Note that, given a
quasi-momentum $k$, the crossings depend on the parameter $\Delta$, and the fidelity will change drastically around
some particular value on $\Delta$, see Fig.~\ref{fig1}. For example, picking $k=q/4$
and $\Delta=0$, the largest overlap with the dressed state with
$\nu=1$ is assumed to be with the $\mu=0$ bare state, assuming a
small coupling $g_0$, while for $-3q^2/4<\Delta<-q^2/4$ the largest
overlap is expected for the $\mu=-1$ bare state.

\begin{figure}[!ht]
\centerline{\includegraphics[width=7cm]{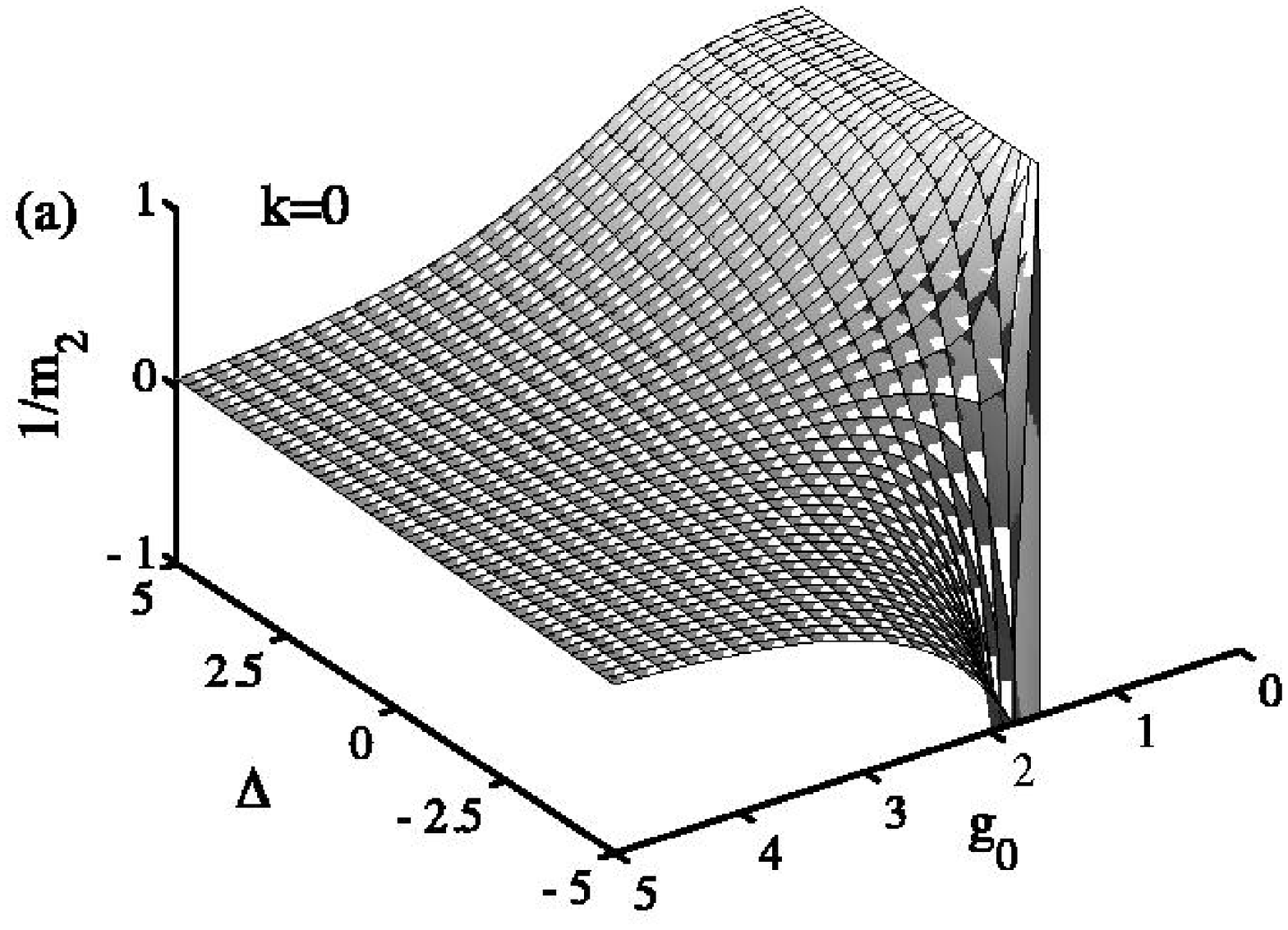}}
\centerline{\includegraphics[width=7cm]{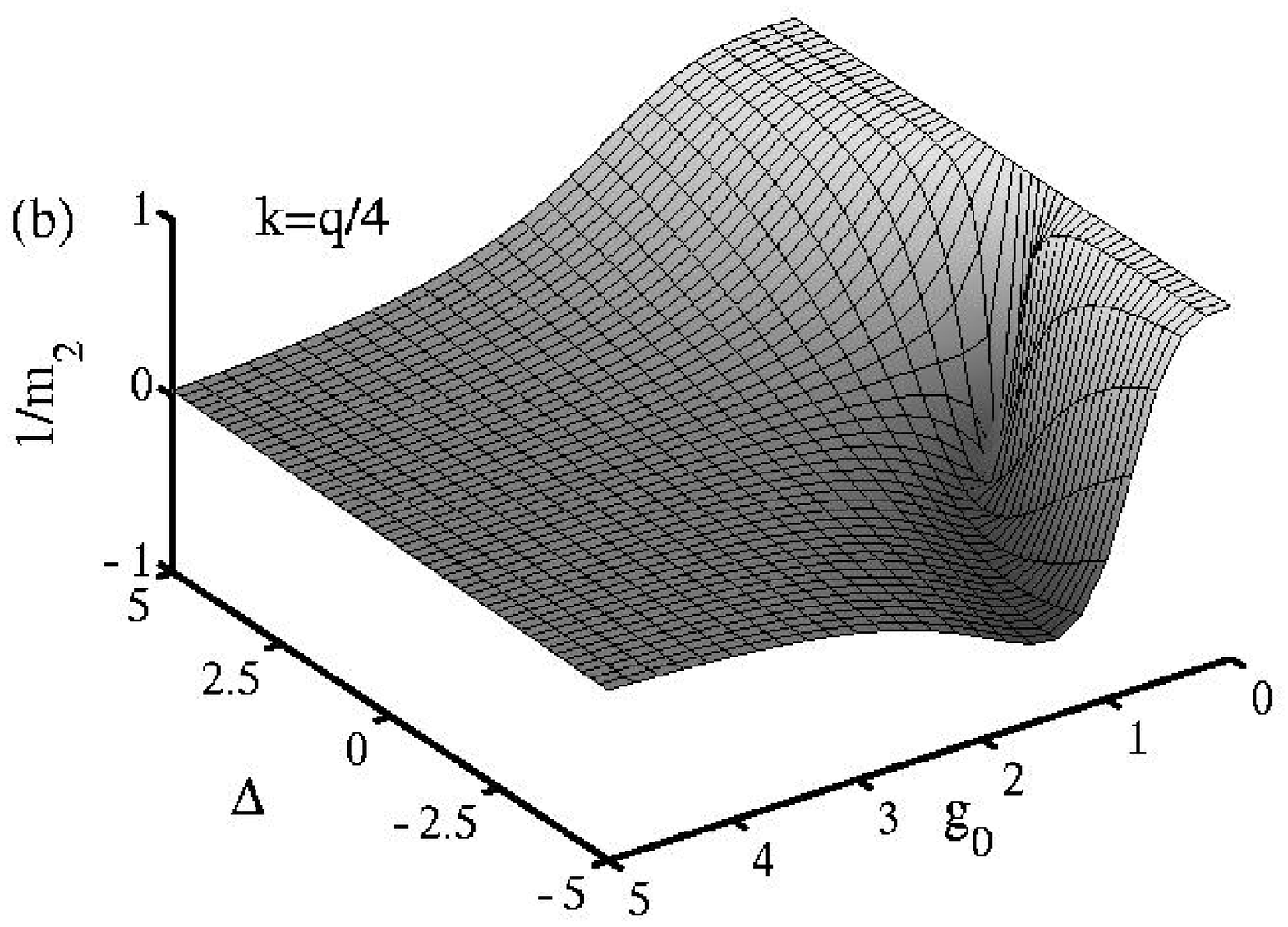}}
\centerline{\includegraphics[width=7cm]{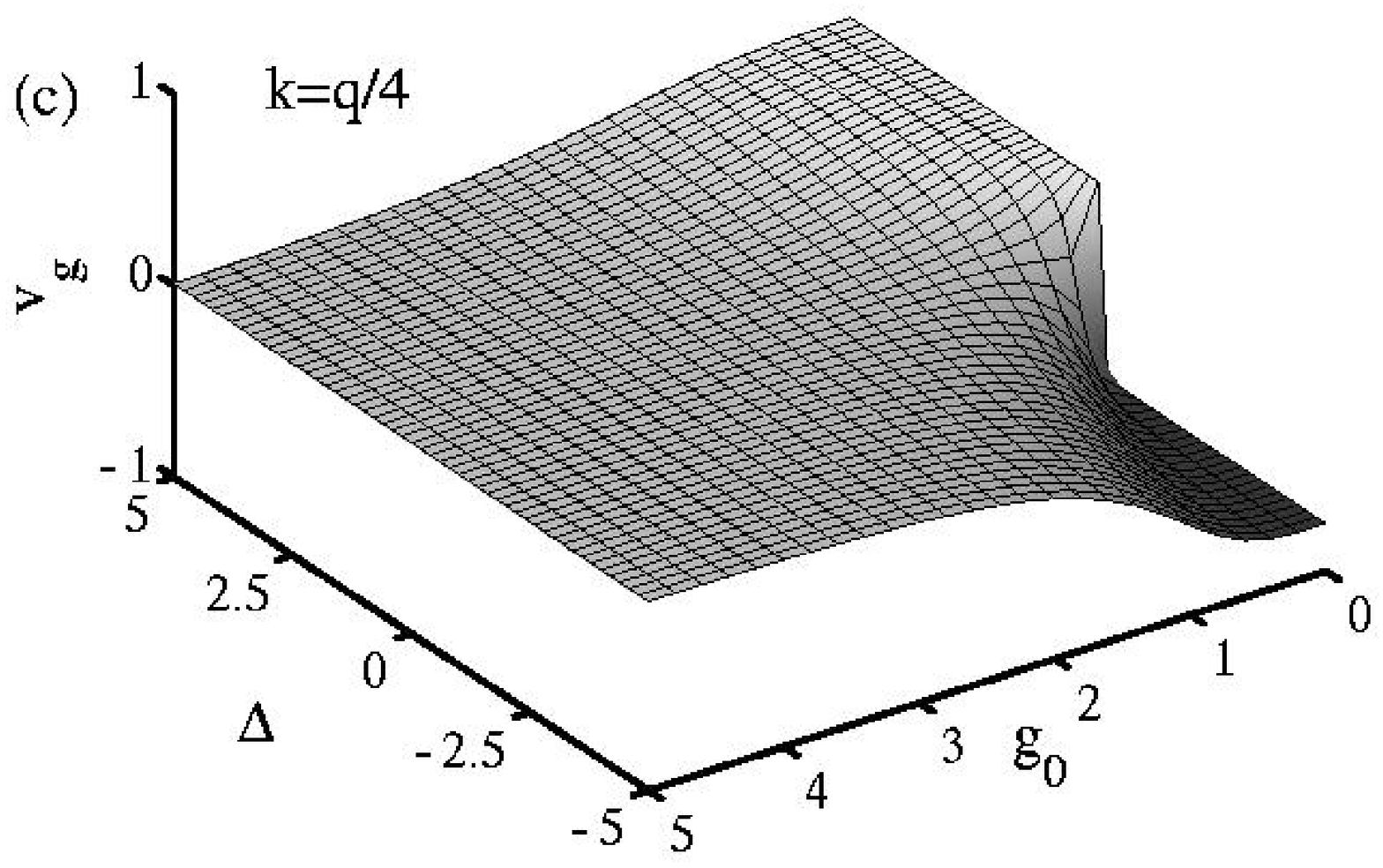}}
\caption[1/(2m^*)]{\label{fig6} The scale mass parameters $1/(m_2)$ and
$v_{{\rm g}}=k_0/m_1$ as functions of $g_0$ and $\Delta$. The first two plots (a) and (b) show
$1/(m_2)$ for $k=0$ and $k=q/4$, respectively, and (c) the group velocity $v_{{\rm g}}$ for $k=q/4$, where the unperturbed
velocity ($g_0=0$) should be $v_0=1/4$. The result derives from
numerical diagonalization of the matrix (\ref{matrix}) and
numerical calculation of the coefficients of the dispersion curve
around the point $k$ of interest. The bare energies of states $\mu=0,-1$ cross for     $\Delta=kq-q^2/2$ and the dominant bare states of the lowest band changes. This leads to a drastic change in the effective parameters, as seen in the figures.}
\end{figure}

In Fig.~\ref{fig6} we display the coefficients $m/(m_2)$ and
$v_{{\rm g}}=k_0/m_1$ as function of $g_0$ and $\Delta$. In
Figs.~\ref{fig6} (a) and (b) $1/(m_2)$ is shown for $k=0$ and
$k_0=1/4$ respectively, and in Fig.~\ref{fig6}(c) the group
velocity is plotted for $k=1/4$.

\begin{figure}[!ht]
\centerline{\includegraphics[width=8cm]{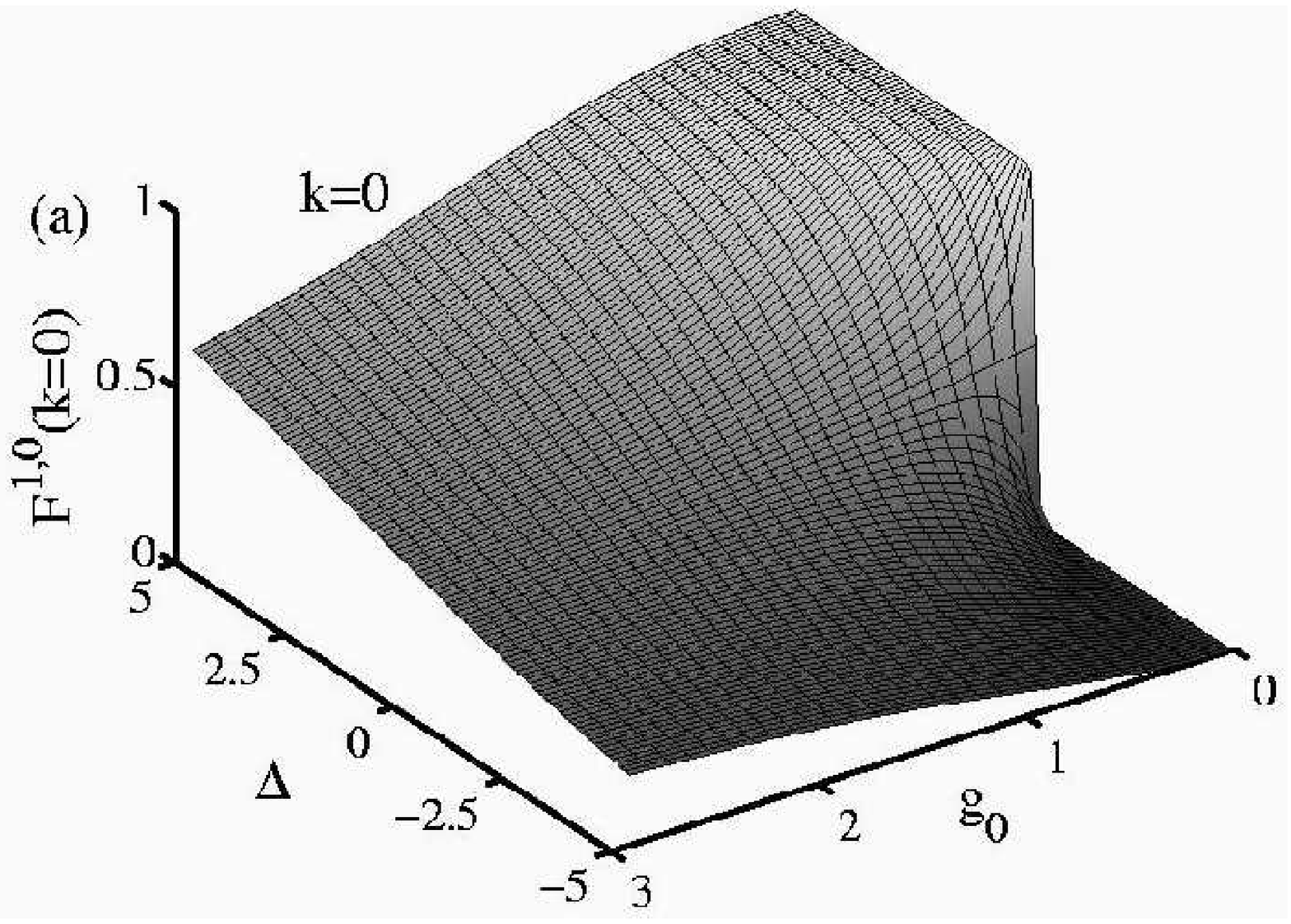}} \vspace{1cm}
\centerline{\includegraphics[width=8cm]{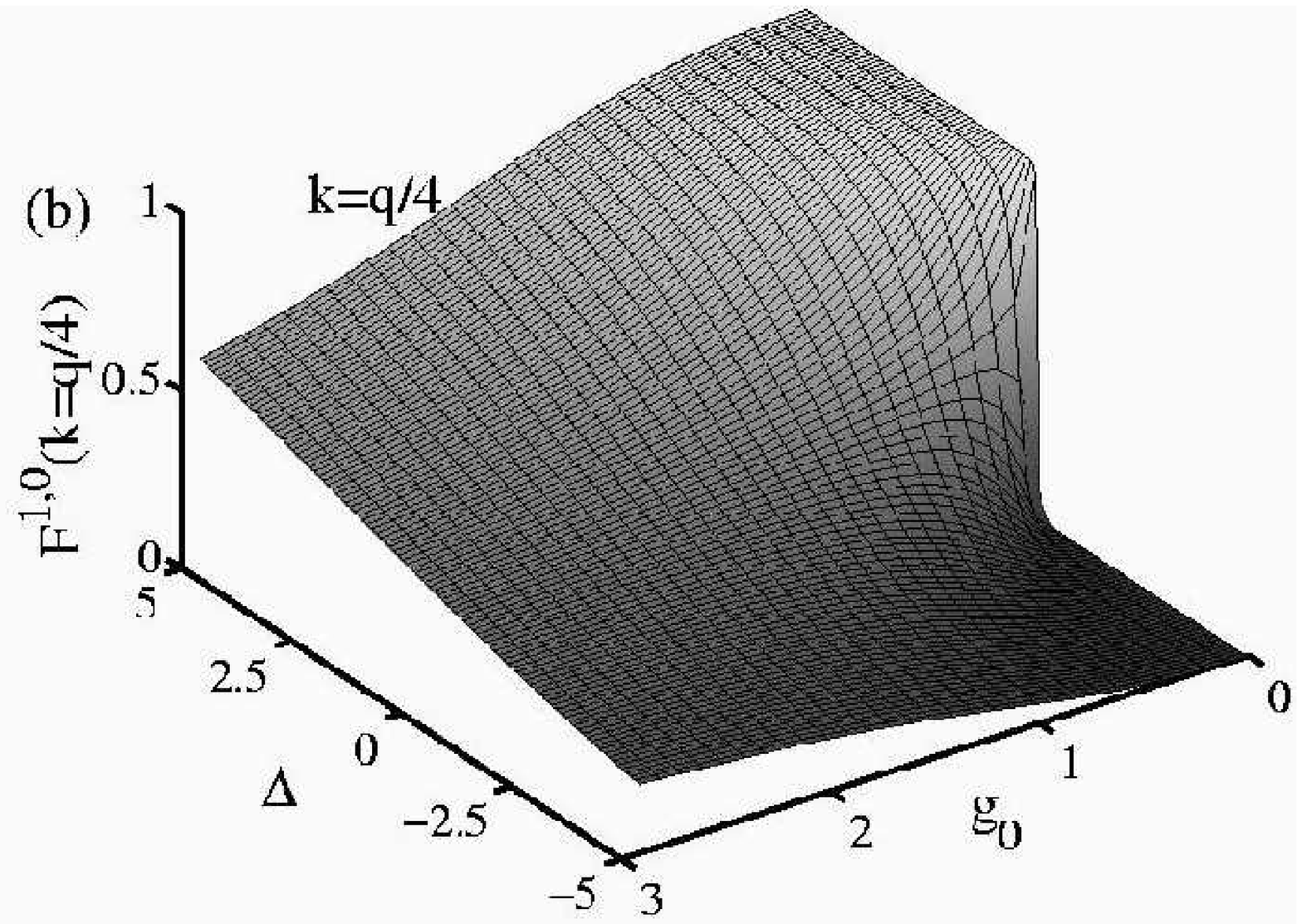}}
\caption[overlap]{\label{fig7} The variation of the overlap (a)
$F^{1,0}(0)$ and (b) $F^{1,0}(q/4)$, as defined in
Eq.~(\ref{overlap}), with respect to $g_0$ and $\Delta=0$. For
positive $k$, a huge change in the overlap occurs when
$\Delta=kq-q^2/2$, which was also seen in the previous figure of the mass parameters. For such $\Delta$, the corresponding
quasi-momentum $k$ is located at a gap, and as $\Delta$ changes
more, the lowest dressed states $\nu=1$ will have a larger overlap
with bare states with $\mu=\pm1$ rather than $\mu=0$. This can be
clearly understood from Fig.~\ref{fig1} (b) and (c). The photon momentum $q$,
which was used to define the characteristic length scale, is, as
earlier 1. }
\end{figure}

From Fig.~\ref{fig6}, it is clear that $1/(m_2)$ differs
considerably from the unperturbed value $m=1$. It is now of
interest to see how well the corresponding bare and dressed states
overlap for the same parameter rangers. We calculate the overlaps
$F^{\nu=1,\mu=0}(k=0)$ and $F^{\nu=1,\mu=0}(k=q/4)$ for the two
examples. The results are plotted in Figs.~\ref{fig7}(a) and (b).
As justified earlier, a small coupling increases the overlap. Note
that there are ranges where the masses $m_1$ (or $v_{{\rm g}}$) and $m_2$
are shifted significantly from the free mass $m$ and still the
overlap is large, meaning that an initial Gaussian bare state
should evolve approximately freely with mass parameters $m_1$ and
$m_2$. In the wave packet simulations below it will be shown that
the shifted masses can be observed also for propagating bare
states.

\subsection{State propagation simulations}\label{ss3c}

Independently of the chosen basis (bare or dressed states), the
time evolution of the atom-cavity system is governed by a
time-independent Hamiltonian $H=p^2/(2m)+V(x)$; with an initial
state $|\Psi(t=0)\rangle$ it is written as
\begin{equation}
|\Psi(t)\rangle={\rm e}^{-{\rm i}Ht}|\Psi(t=0)\rangle.
\end{equation}
For an $x$-dependent potential $V(x)$, we find, in general, that
the kinetic energy term of the Hamiltonian does not commute with
the potential. The exponential can therefore not be separated into
one momentum exponential and one spatial exponential. However, if
the time of propagation $\delta t$ is chosen sufficiently small,
the error of splitting up the exponential becomes negligible
\begin{equation}
|\Psi(t+\delta t)\rangle\approx {\rm e}^{-{\rm i}\frac{p^2}{2m}\delta t}{\rm e}^{-{\rm i}V(x)\delta t}|\Psi(t)\rangle.
\end{equation}
This is called {\it the  split operator method}, see
Ref.~\cite{split}, and the procedure goes as follows: Starting
with some state $|\Psi(0)\rangle$ in the $x$-representation, we
multiply it by the exponential $\exp(-{\rm i}V(x)\delta t)$, we then
take the Fourier transform of the state and obtain it in the
$p$-representation, and then we multiply it by
$\exp\left[-{\rm i}p^2\delta t/(2m)\right]$, and finally transform it
back to the $x$-representation with the inverse Fourier transform,
which gives us $|\Psi(\delta t)\rangle$. Repeating this $N$ times
we obtain $|\Psi(t)\rangle$, for $t=N\delta t$.

For the system considered here, the kinetic and potential terms are
\begin{equation}
\begin{array}{ccc}
E_{\rm kin} &= \frac{p^2}{2}\left[\begin{array}{cc} 1 & 0 \\ 0 & 1\end{array}\right]\\ \\
V(x)         &= \left[\begin{array}{cc} \frac{\Delta}{2} & g(x) \\ g(x) & -\frac{\Delta}{2}\end{array}\right].
\end{array}
\end{equation}
If the atom is initially in a bare state corresponding to the internal state $|-\rangle$, the wave function is
\begin{equation}
|\Psi(t=0)\rangle=\left[\begin{array}{c}0 \\ \int dk\,\varphi(k)|k\rangle\end{array}\right]
\end{equation}
while, for an initial dressed state, it is given by Eq.(\ref{dressed_Gaussian}) for $t=0$.

\subsection{Propagation within the cavity}

For a given $k_0$, the properties of the dispersion curve, and
hence of the mass parameters, depend on the physical parameters of
the system. From Eq.~(\ref{appresult2}), it is
expected that effects are largest for $\Delta=0$ as long as $k_0$
is far from a crossing, and thus we choose $\Delta=0$ in the
analysis below. Furthermore, we choose to calculate the quantities
for $k_0=0$ and $k_0=q/4$ in the lowest band $\nu=1$.

\begin{figure}[!ht]
\centerline{\includegraphics[width=8cm]{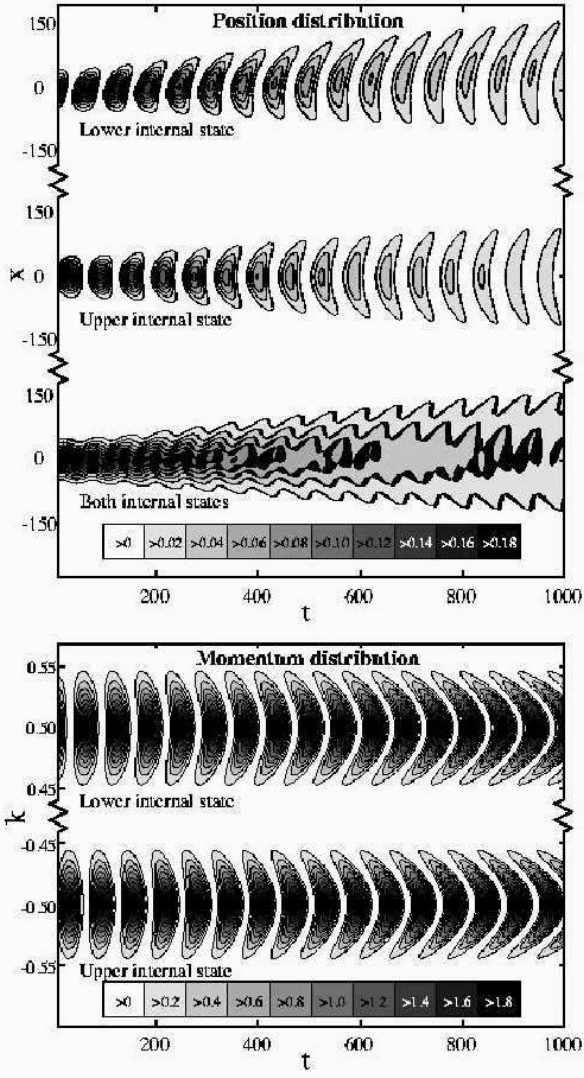}}
\caption[forbidden]{\label{forbidden} The evolution of an initial
Gaussian bare wave packet prepared inside the cavity with
$\Delta_x^2=500$ and $k_0=1/2$, corresponding to a quasi-momentum of
the first forbidden energy gap. The upper plot shows the wave
packets of the ground, excited and total (ground plus excited)
atomic state in the $x$-representation. The lower figure displays the
wave packet of the ground and excited atomic state in the momentum
$k$-representation. It is clear that we have a spreading in $x$
and also a small drift of the lower atomic wave packet. The population 'Rabi'
oscillates between the two states with a period
$T_R\approx\pi/g_0$, corresponding to the $2\pi$ divided by the
energy gap size $2g_0$ in first order. The remaining parameters
are $\Delta=0$ and $g_0=0.05$. The contour-bars gives the relative values.    }
\end{figure}

We will, in this Subsection, show numerical simulations and
calculations of the effective parameters for a Gaussian wave
packet inside a cavity, but first we investigate the evolution of
a bare Gaussian wave packet that is prepared inside the cavity
with a momentum $k_0=q/2$, corresponding to a gap. At an energy gap,
the dispersion curve has a vanishing first order derivative. Hence, if the wave packet has a large overlap with the
corresponding dressed states, its group velocity is close to zero. In
the weak coupling limit, the time dependent bare state can be
expressed in terms of dressed states approximately as
\begin{equation}\begin{array}{ccl}
|\psi(t,k_0=0.5)\rangle& \approx & \frac{1}{\sqrt{2}}\left({\rm e}^{-{\rm i}E^{\nu=1}t}|\phi_1(k_0=q/2)\rangle\right. \\ \\ & & +\left.{\rm e}^{-{\rm i}E^{\nu=2}t}|\phi_2(k_0=q/2)\rangle\right),
\end{array}
\end{equation}
and using that the gap size is approximately $2g_0$ it follows
\begin{equation}
\begin{array}{ccl}
|\psi(t,k_0=q/2)\rangle& \approx& \frac{1}{\sqrt{2}}\left({\rm e}^{i{\rm }g_0t}|\phi_1(k_0=q/2)\rangle\right. \\ \\ & & +\left.{\rm e}^{-{\rm i}g_0t}|\phi_2(k_0=q/2)\rangle\right){\rm e}^{-{\rm i}\mathcal{E}^0t}.
\end{array}
\end{equation}
Thus, the bare state wave packet is supposed to oscillate between
upper and lower states $|\pm\rangle$. Figure \ref{forbidden} shows
the evolution of such a prepared state. The upper figure shows the
wave packets in the $x$-representation for upper, lower and combined internal states, while the lower plot displays the momentum packets.

\begin{figure}[ht]
\centerline{\includegraphics[width=8cm]{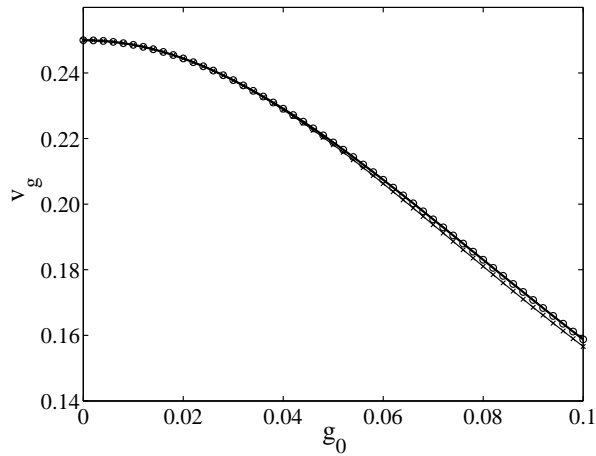}}
\caption[effmass]{\label{fig14} The group velocity $v_{{\rm g}}$ as
function of the coupling, obtained, both from the wave packet
simulation, circles shows the result from bare state propagation
and crosses from dressed state propagation, and the Floquet theory
(solid line), when $k_0=q/4$. A clear agreement is seen, but
unexpectedly the bare state result coincide better with the
Floquet one, which might be due to a not fully adiabatic
preparation of the dressed state. The parameters are
$t_{\rm f}=1000$, $\Delta=0$ and $\Delta_x^2=2500$. }
\end{figure}

When the atom propagates inside an infinite cavity, it should be
possible to describe the atomic wave packet as a freely evolving
particle with effective mass parameters defined in
Eq.~(\ref{E_approx}), which will be analyzed next. The time of
flight $t_{\rm f}$ between $x_0$ and $x_{\rm f}$ is given by
\begin{equation}
t_{\rm f}=\frac{x_{\rm f}-x_0}{v_{{\rm g}}}=\frac{m_1}{m}\frac{(x_{\rm f}-x_0)}{v_{\rm initial}},
\end{equation}
where $v_{\rm initial}$ is the velocity of the atom in absence of
the cavity interaction. We will simulate propagation of both bare and dressed
wave packets. In the simulation of bare state propagation we use
averages, for the atomic state of interest, that is for the wave
packet of the ground state, $\langle x\rangle=\frac{\langle-|x|-\rangle}{\langle-|-\rangle}$, in other words we select the wave
packet of the ground state $|-\rangle$ and normalize it and then
calculate the averages. Physically that assumes that a
projective measurement is carried out on the lower state before
that position measurement is done. For the dressed states the
averages are calculated for the whole atomic wave packet and no
projective measurement is assumed. We propagate the initial wave
packet from a certain $\langle x_0\rangle$ for a time $t_{\rm f}$
and calculate the final average position $\langle
x_{\rm f}\rangle$, and deduce the propagation velocity
$v=\frac{\langle x_{{\rm f}}\rangle-\langle x_0\rangle}{t_{{\rm f}}}$, which
should coincide with the group velocity $v_{{\rm g}}=k_0/m_1$.  We use the
parameters of one of the previous examples, $\Delta=0$,
$\Delta_x^2=2500$ and $k_0=q/4$, but vary the coupling strength
$g_0$; the final time is taken to be $t_{\rm f}=1000$. During the
adiabatic preparation, the dressed wave packet will already
broaden, whence its initial width is no longer $\Delta_x^2=2500$,
but this is not expected to affect the group velocity. In
Fig.~\ref{fig14} we compare propagation velocities for dressed and
bare states with the group velocity obtained from the Floquet theory.
The three curves coincide rather well, but, for the case of initial bare state, the agreement is good only for the lower initial state $|-\rangle$, as described above. The whole atomic wave packet, containing
both upper and lower atomic states, gives a value $\langle
x_{{\rm f}}\rangle$ different from the one using just one atomic state.
Thus, we emphasize again that, extracting the propagation velocity
with the bare states assumes that the position measurement is
carried out for the atom in its lower state, in other words, some
of the outcomes will fail, namely when the atom is measured while
being excited. The slight difference between dressed state
propagation and the Floquet theory might be due to non-adiabatic
generation of the initial dressed state.

This scheme clearly gives a
good measure of the group velocity and from that it is possible to
calculate the mass $m_1$.

\begin{figure}[ht]
\centerline{\includegraphics[width=8cm]{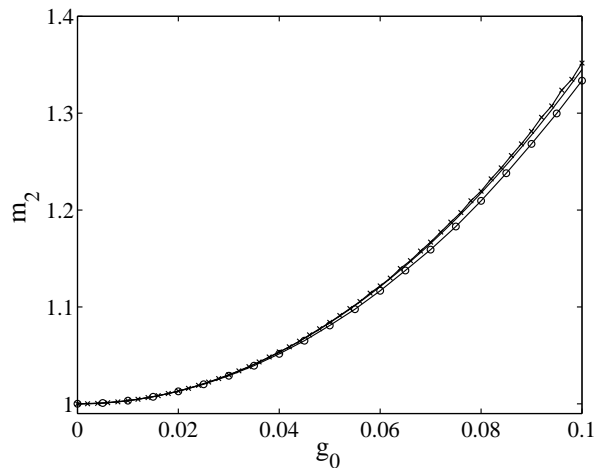}}
\caption[effmass2]{\label{fig15} The effective mass $m_2$ as
function of the coupling, obtained, both from the wave packet
simulation of bare (circles) and dressed states (crosses) and the
Floquet theory (solid line), when $k=0$. The main reason for the
deviation of the  result of the bare state propagation is that the wave
packet of the lower atomic state, which is used in the
calculation, does not stay Gaussian and the tails of the
distribution are cut off for $|x|>15$ in the calculation of
$\Delta_x^2$. The parameters are $t_{\rm f}=1000$, $\Delta=0$ and
$\Delta_x^2=1500$.  }
\end{figure}

The mass parameter $m_2$, enters the wave
packet dynamics through the broadening of the Gaussian dressed state 
according to
\begin{equation}\label{dressbroaden}
\Delta_x^2(t)=\langle x^2\rangle=\frac{1}{4\Delta_k^2}+\left(\frac{\Delta_k}{m_2}\right)^2t^2.
\end{equation}
Initial Gaussian bare states do not, however, remain Guassian but split into several wave packets, corresponding to a number of dressed states. In Fig.~\ref{gaussfig}
(b) the initial bare state splits into three main wave packets and,
even though the two outer ones may be small, they contribute
significantly to the variance since they are far from the center
packet. Thus, for the calculation of the bare state variance, the
packets in the tails $|x|>150$, which mostly corresponds to excited
atomic states, are excluded when $\langle x^2\rangle$ is
calculated. In Fig.~\ref{fig15} the mass $m_2$ is displayed as
function of $g_0$. The time of 
propagation is $t_{{\rm f}}=10$, and the mass is obtained by a least
square fit of Eq.~(\ref{dressbroaden}), knowing $\Delta_x^2(t)$
and $t$. Note that when we prepare the Gaussian dressed states by
an adiabatic turn on of the coupling amplitude,
Eq.~(\ref{dressbroaden}) is valid, but the mass $m_2$ is not
constant during the turn-on, which must be taken into account.
This is carried out by assuming a constant coupling amplitude, adjusted
by introducing an effective turn-on time $t_{{\rm eff}}$. Thus
Eq.~(\ref{dressbroaden}) becomes
\begin{equation}
\Delta_x^2(t)=\frac{1}{4\Delta_k^2}+\left(\frac{\Delta_k}{m_2}\right)^2(t_{{\rm eff}}+t)^2,
\end{equation}
where the mass $m_2$ is now constant during the turn-on process.
In lowest order, the mass $m_2$ behaves as $\propto g_0^2$, according to
Eq.~(\ref{appresult2}), which is indicated in Fig.~\ref{fig15}
and it has been confirmed by a polynomial fit to the data, where the
dominant coefficient is the one of $g_0^2$.

The measurements of the masses proposed in this Subsection may be
used for state preparation. The masses are functions of $g_0$,
$\Delta$ and $q$ for the one excitation case, but in general it is
a function of the effective coupling $g_0\sqrt{n}$, where $n$ is
the photon number, instead of $g_0$. Thus, for a general initial
state of the cavity field, a perfect mass measurement will give
$m_{1,2}=m_{1,2}(g_0\sqrt{n},\Delta,q)$ for any $n$. Only masses
$m_{1,2}(g_0\sqrt{n},\Delta,q)$ with $n$'s that has a
non-vanishing probability from the initial photon distribution,
can be detected. Knowing $m_1$ or $m_2$ and $g_0$, $\Delta$ and
$q$ it is possible to solve for $n$, meaning that measuring the
mass reduces the field to the Fock state $|n\rangle$, a sort of
projective measurement. Physically this means that an initial
state $|\Phi(t=0)\rangle$ splits up in a set of states
$|\Phi(t)\rangle_n$ for the various photon numbers $n$. If all
overlaps between different final states $|\Phi(t)\rangle_n$ vanish
it is possible to separate the individual parts with one single
measurement. However, having non-overlapping final states seems
unlikely in realistic experiments.

\subsection{Scattering and transmission}

In order to simulate atomic scattering and transmission by the
cavity, the mode function coupling $g(x)$ needs to be modified by multiplication of 
an envelope function
\begin{equation}\label{envelope}
\bar{g}=\frac{1}{2}\left[\tanh\left(\frac{x+x_{{\rm l}}}{x_{{\rm e}}}\right)-\tanh\left(\frac{x-x_{{\rm l}}}{x_{{\rm e}}}\right)\right].
\end{equation}
This function goes to zero for large $|x|$ and is centered around
$x=0$, the cavity length is given by $2x_{{\rm l}}$, and the 'slope' how
fast it turns on/off is determined by $x_{{\rm e}}$. Thus, the coupling
will approximately be zero for $x<-x_{{\rm l}}$ and $x>x_{{\rm l}}$, and in the
interval $[-x_{{\rm l}},x_{{\rm l}}]$ it behaves as $g(x)=2g_0\cos(qx)$. We choose
$x_{{\rm l}}$ such that $2x_{{\rm l}}\ll\lambda=2\pi/q$,
and $x_{{\rm e}}$ such that the coupling is turned on/off smoothly, but
fast enough for a non-adiabatic transition. The envelope function allows us to test boundary effects
from not having a completely periodic potential; it turns out,
however, that the dynamics follows very well the Floquet
predictions as long as $\lambda\ll 2x_{{\rm l}}$.

An incoming atom entering a cavity has been studied in a series of
papers~\cite{refl1,refl2,refl3}, were, in particular, the
transmission and reflection coefficients have been calculated. In
these papers, either a mesa function or a hyperbolic secant
squared coupling $g(x)$ was used to obtain analytical solutions.
The spatial wave function was assumed to be a plane wave and not a
propagating wave packet. In the case of a mesa function or a
hyperbolic secant, the atom effectively sees a potential barrier
or a well with an amplitude $\pm\sqrt{\Delta^2/4+g^2(x)n}$, where
$n$ is the photon number. The case of a standing wave shape of the
coupling is, however, different; the spectrum of allowed energies is then
formed by bands, with forbidden gaps in between. Thus, an atom
entering the cavity mode must have an allowed energy in order to
traverse the cavity. If the atom has an energy falling into
forbidden energy gaps, it must be reflected (assuming the tunneling rate through the cavity to be negligible).

Going back to Fig.~\ref{fig1}, we see that, when $\Delta=0$, the
crossings occur for $(\mu+1/2)q$ and $\mu q$, where, as before,
$\mu=0,\pm1,\pm2,...\,$ . We know that these become avoided
crossings when the coupling is turned on, and they give the
forbidden energy gaps in the spectrum. The gap size decreases for
increasing band index $\nu$, and the largest gap is at $k=\pm q/2$
between the first two bands. If the momentum distribution
$\varphi(k)$ of the incident atom is such that $\langle
k\rangle\approx0.5q$ and the energy spread, due to the spread
$\Delta k$, is smaller than the gap width, the atom should be
reflected when it approaches the cavity mode, and from energy
conservation we must have $\langle
k\rangle\approx0.5q\,\Rightarrow\langle k\rangle\approx-0.5q$. The
atom thus shifts the momentum by one unit $q$ and, due to the form
of the Hamiltonian (\ref{ham1}), also the internal atomic-cavity
state flips. Hence the quantum state is changed, by the
reflection against the lowest gap, as
\begin{equation}
\int\,dk\,\varphi(k)|k\rangle|\pm\rangle\,\,\Rightarrow\,\,\int\,dk\,\varphi(k)|k-q\rangle|\mp\rangle.
\end{equation}
On the other hand, the internal atomic state will not flip for
scattering against a gap corresponding to momentum $\mu q$. If the
atom goes from $|\uparrow\rangle$ ($|\downarrow\rangle$) to
$|\downarrow\rangle$ ($|\uparrow\rangle$), the photon number must also change
as $n\rightarrow n\pm1$. This means, for example, that starting
with the vacuum and reflecting $N$ excited atoms from the cavity
allows one to create an $N$-photon Fock state in the mode. If instead
the atoms scatter when they are initially in their ground state,
the mode will be cooled down towards lower photon numbers.

\begin{figure}[!htb]
\centerline{\includegraphics[width=8cm]{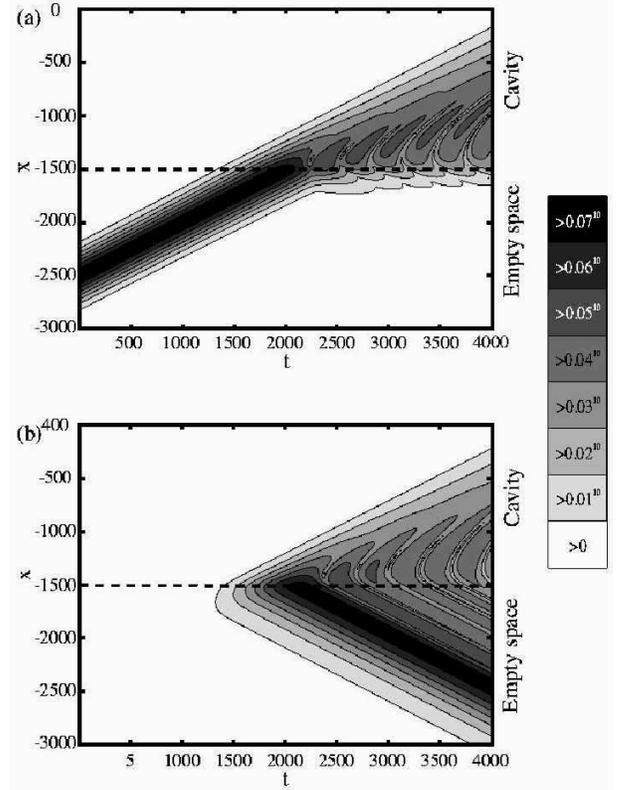}}
\caption[reflection1]{\label{fig8} Scattering of an atom against a
cavity. The cavity is located within $|x|\lesssim x_{{\rm l}}$ for
$x_{{\rm l}}=1500$ (marked with dashed line) and $x_{{\rm e}}=50$. In (a) we show
the ground state atomic wave function $|\langle-|\langle
x|\Psi\rangle|^2$ and in (b) the upper state $|\langle+|\langle
x|\Psi\rangle|^2$ as function of position $x$ and time $t$. Note
that the atom starts to propagate in its lower state at $x_0=-2500$
at the time $t=0$ with momentum $k_0=q/2$ until it reaches the
cavity boundary at $x=-1500$. It is reflected and its internal
state is flipped to the upper state $|\uparrow\rangle$. The
reflection is due to the fact that its energy coincides with an
energy gap; hence the atom absorbs the photon from the cavity
field and the impact changes the direction of propagation. Note
the scale of the contours, which indicate that almost everything is
reflected, as confirmed in Fig.~\ref{fig11} for the atomic
inversion. The parameters are here are $g_0=0.01$, $\Delta=0$, and
$\Delta_x^2=2500$. The contour-bar shows relative values.}
\end{figure}

Figure~\ref{fig8} shows the results of a numerical simulation, in
which an atom reflects from a cavity. The atom is initially in its
lower atomic state $|-\rangle$ and ends up in $|+\rangle$. The
cavity starts at around $x=-150$ and ends at $x=150$, $x_{{\rm l}}=1500$
and $x_{{\rm e}}=50$, and the flip takes place at the edge of the cavity.
The scale of the contours is chosen such that even the
small amplitudes of the wave packet is seen; a closer look on the
scale indicates that almost all the population is reflected. 

\begin{figure}[ht]
\centerline{\includegraphics[width=8cm]{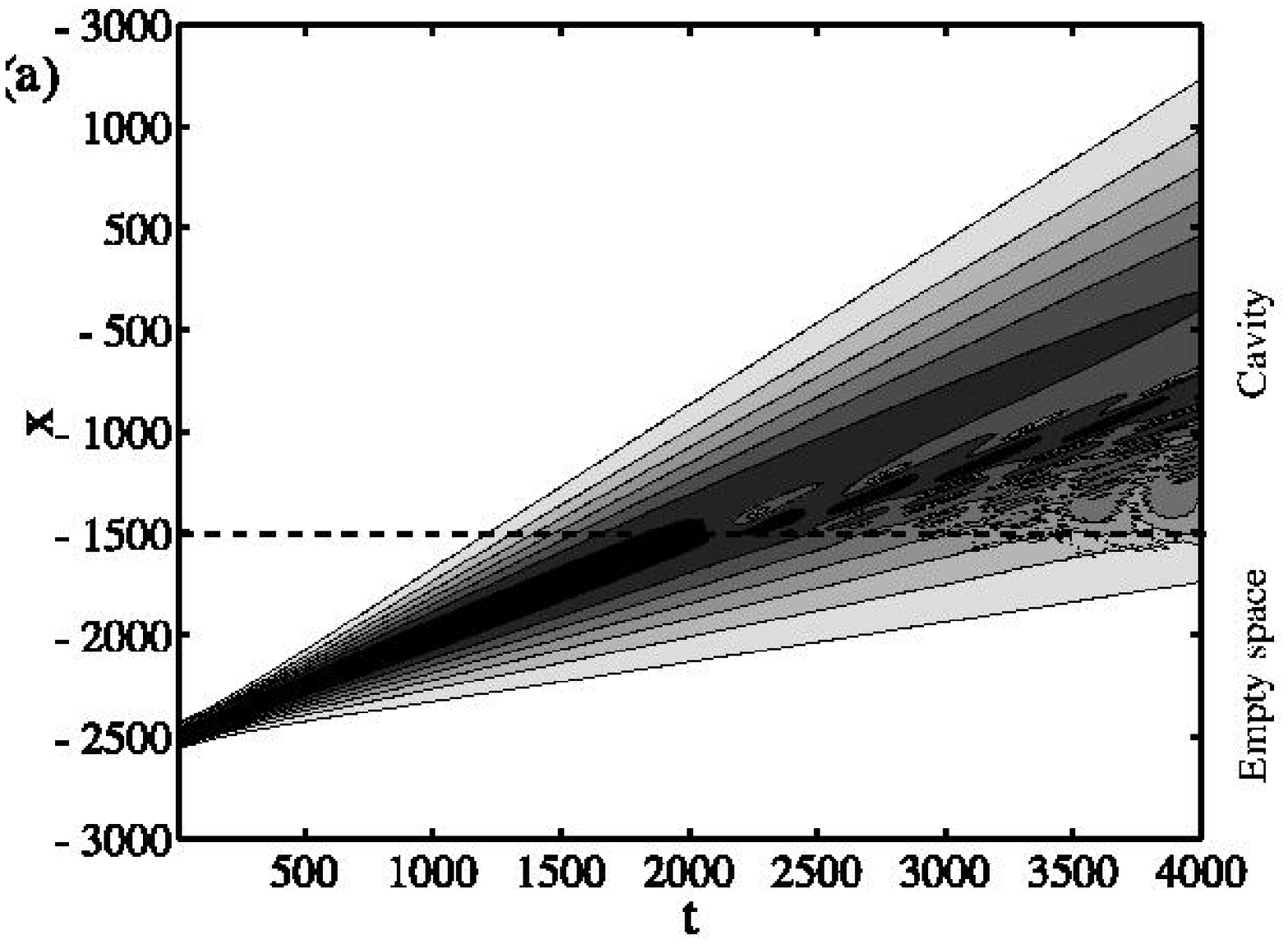}} \vspace{0cm}
\centerline{\includegraphics[width=8cm]{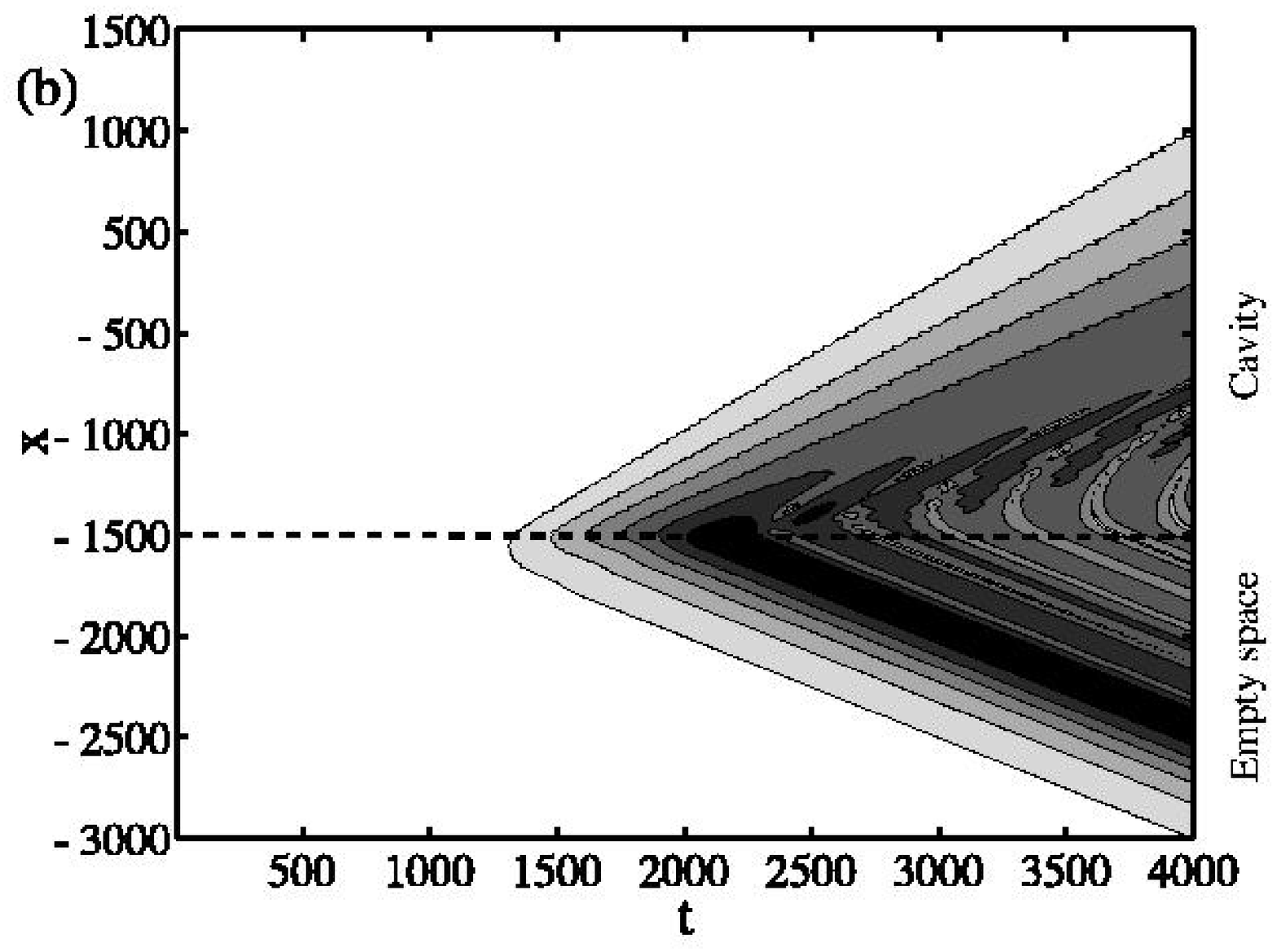}}
\caption[momreflection2]{\label{fig9} This figure shows the same
as Figs.~\ref{fig8} (a) and (b), except that now the momentum
distribution is much wider, $\Delta_x^2$ is chosen 100 instead of 2500
as in the previous plot. The consequence is that energy spread
exceeds the band gap and the tails of the momentum wave packet are
transmitted rather than reflected. The contour scale is the same
as in Fig.~\ref{fig8}, and a closer look shows that a much larger
part of the wave packet is propagating inside the cavity. Note how the wave packets spreads much
faster in these plots. }
\end{figure}

\begin{figure}[ht]
\centerline{\includegraphics[width=8cm]{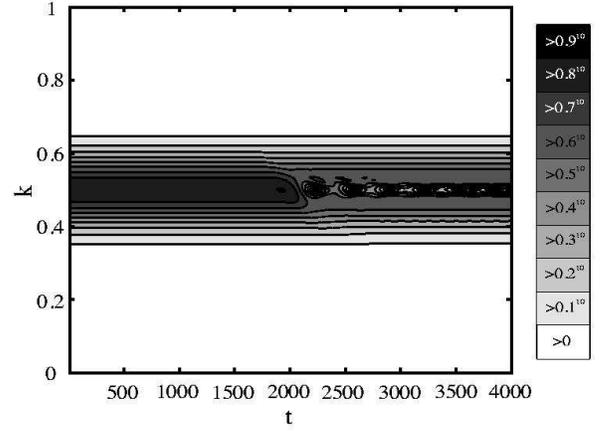}} \vspace{0cm}
\caption[momreflection2]{\label{fig10} The momentum distribution
of the lower atomic state, $|\langle-|\langle k|\Psi\rangle|^2$,
corresponding to Fig.~\ref{fig9}, as a function of time $t$. The
forbidden energies are 'burned out' when the packet hits the
cavity.  The contour-bar gives relative values.}
\end{figure}

If the energy spread of the incident atom is broader than the band
gap, the tails of the packet are in the allowed region of
energies, and can therefore enter the cavity. Figure~\ref{fig9}
depicts the same system as Fig.~\ref{fig8}, but the
momentum distribution is much wider, hence part of the wave packet
is transmitted. This phenomenon can also be observed
on the momentum representation of the wave packet: The part of the
wave packet having allowed energies is transmitted, while the
forbidden part is reflected, with a 
flipped atomic state $|\downarrow\rangle\rightarrow|\uparrow\rangle$.
Thus, due to the forbidden energies, a 'hole' is seen in the
momentum distribution. This suggests a way of measuring the
band/gap structure, since the energies of the reflected atoms
correspond to gaps. The momentum wave packet for
the $|-\rangle$ atomic state is shown in
Fig.~\ref{fig10}. We observe that the momenta close to
$k=q/2$ are 'burned out'. 

\begin{figure}[ht]
\centerline{\includegraphics[width=8cm]{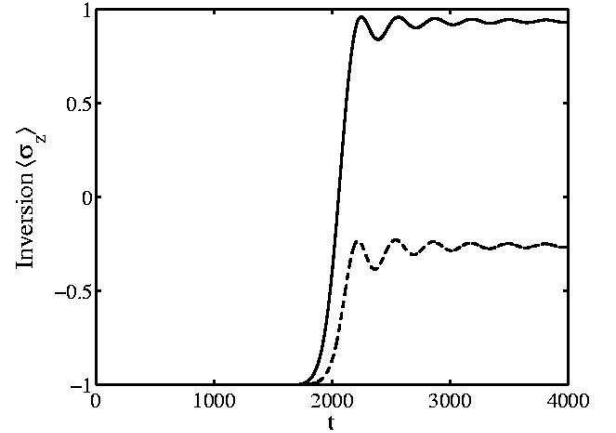}}
\caption[inversion]{\label{fig11} The atomic inversion
$\langle\sigma_3\rangle$ as function of time $t$ for the example
in the previous Fig.~\ref{fig8} (solid line), and Fig.~\ref{fig9}
(dashed line). It is clear that at the cavity boundary, the atom
flips from $|\downarrow\rangle$ to $|\uparrow\rangle$ in the first
example, but when the momentum width becomes larger than the gap,
parts of the wave packet will not be reflected and therefore will
not be flipped.}
\end{figure}

The atomic inversion, defined as $\langle\sigma_3\rangle$, indicates the amount of the atomic population that is in the upper
respectively the lower state. If $\langle\sigma_3\rangle=-1$, all
the population is in the lower state, while if
$\langle\sigma_3\rangle=1$, the atom is completely excited. In
Fig.~\ref{fig11} we show the inversion for the two examples above,
$\Delta_x^2=2500$ and $\Delta_x^2=100$, as function of time. Here it
is obvious that, in the first case, almost the entire 
population is flipped, while in the second most of it is not
flipped. We note that, after the atomic wave packet has reached
the cavity, an oscillating behavior can be observed. The reason is
the interference between the dressed states involved.

\begin{figure}[ht]
\centerline{\includegraphics[width=7cm]{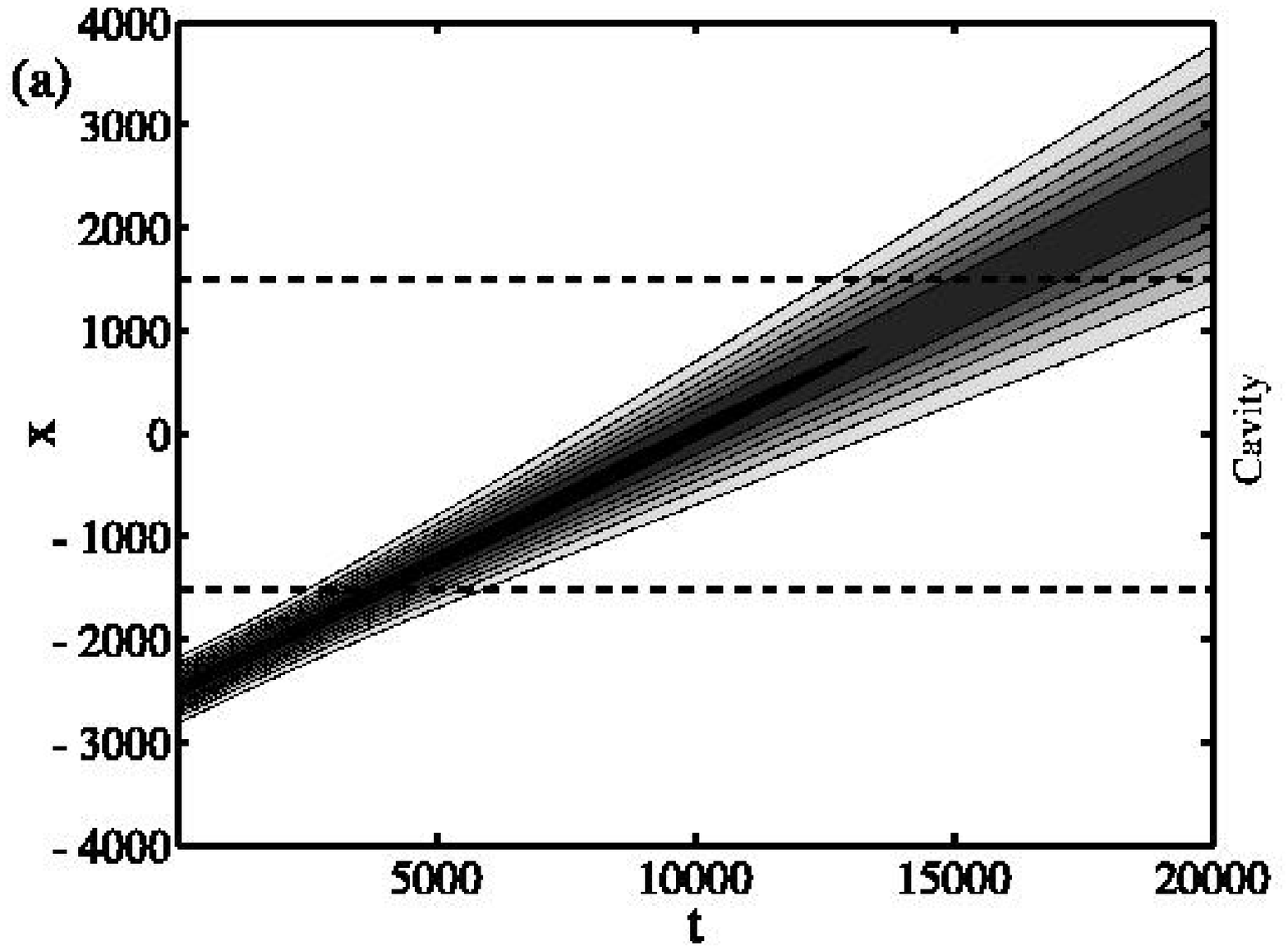}} \vspace{0cm}
\centerline{\includegraphics[width=7cm]{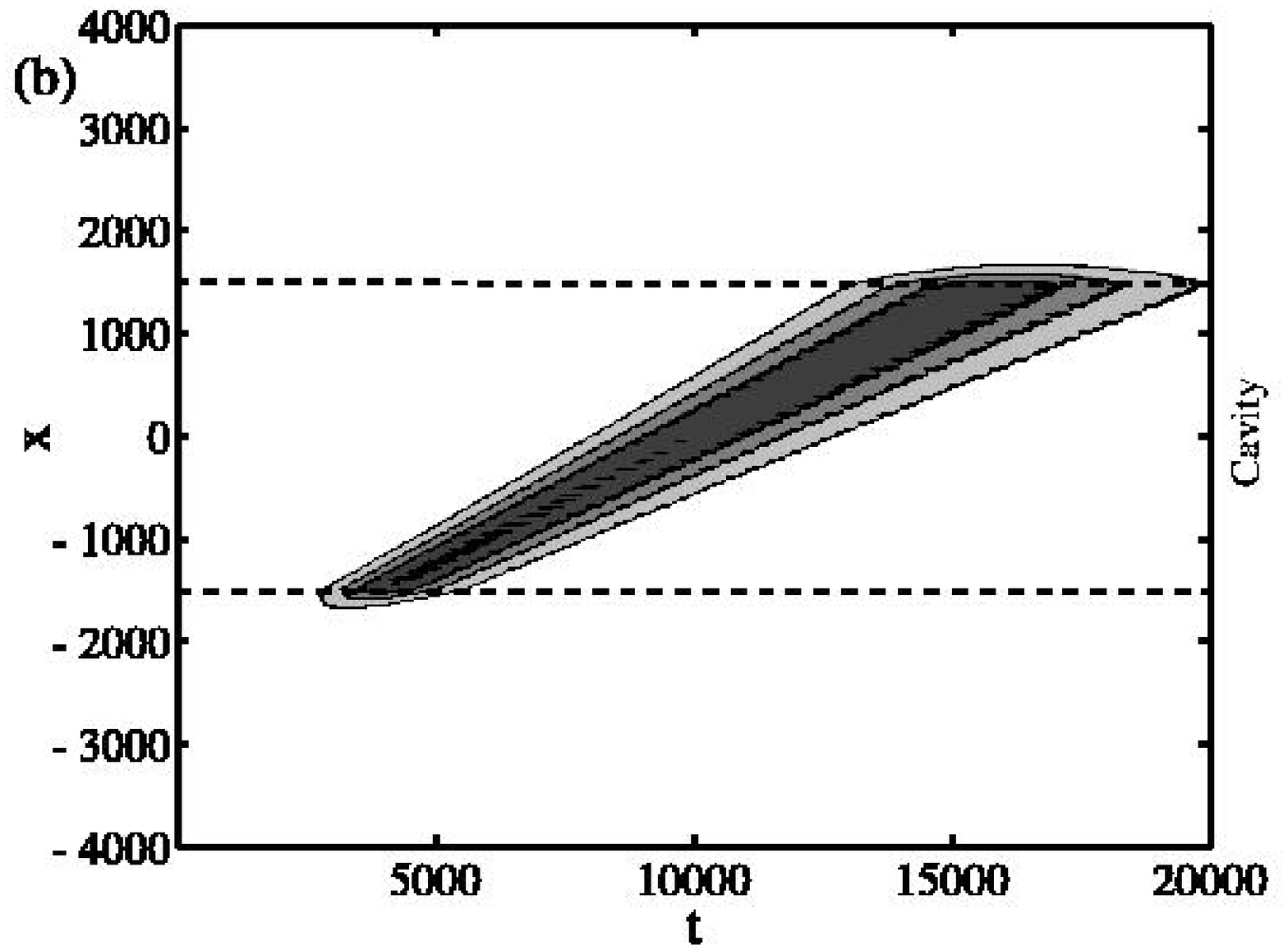}}
\caption[transmit]{\label{fig12} Transmission of the atom through
the cavity. The plots are the same as in Figs.~\ref{fig8}(a) and
(b), but now the initial momentum is $k_0=q/4$, and the atom
energy falls on the first allowed energy band, contrary to Figs.~\ref{fig8} and \ref{fig9}. The overlap between the initial bare
state $|\Psi_0^{k_0=1/4}\rangle$ and the dressed state
$|\phi_1^{k_0=1/4}\rangle$ is 0.998. Thus, the propagation can be
described using the effective parameters and the wave packet
remains Gaussian; only a small amount of population is transferred
into the $|+\rangle$-state and the Gaussian wave packet spreads
according to the effective mass $m_2$ and traverse the cavity with
a velocity $v_{{\rm g}}$. The contour scale is the same as in
Fig.~\ref{fig8} and the cavity boundaries are marked by the dashed
lines.}
\end{figure}

\begin{figure}[ht]
\centerline{\includegraphics[width=8cm]{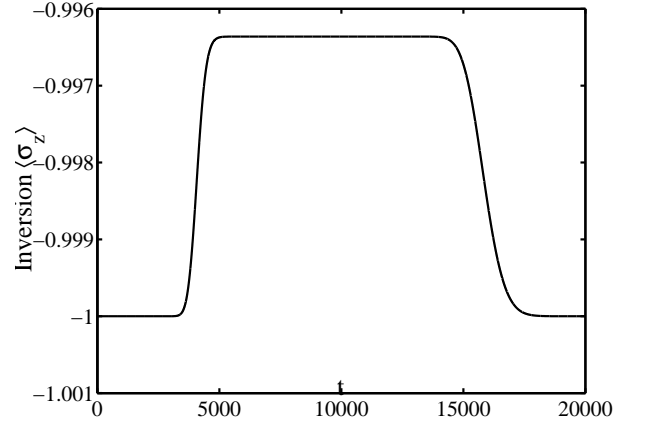}}
\caption[inversion2]{\label{fig13} The atomic inversion
$\langle\sigma_z\rangle$, corresponding to Fig.~\ref{fig12}. Note
that when the atom has traversed the cavity, it regains its
original state. The asymmetry is due to broadening of the wave packet while traversing the cavity, leading to a longer transit time as it leaves the cavity compared to when it enters.
}
\end{figure}

So far the atom has entered the cavity with a momentum $\langle
k\rangle$ that lies within  a band gap, but we now consider the
situation in which it is within the first allowed band $\nu=1$. We still choose the
atom to start in its ground state $|-\rangle$, but now with
$k_0=q/4$, halfway between $k=0$ and the gap at $k=0.5q$. If the overlap between the
initial bare state and the dressed state $|\phi^k_1\rangle$ is close to
unity, the population of the internal states of the system stays approximately
constant, and the atom will traverse the cavity without changing
between its two states $|\uparrow\rangle$ and $|\downarrow\rangle$ nor change its
momentum state. It will, however, experience a shift in the masses
$m_1$ and $m_2$. The overlap for these parameters is
$F^{\nu=1,\mu=0}(k_0=1/4)=0.998$, which is close to 1. The result
of the simulation is given in Fig.~\ref{fig12}.  The atomic
inversion is shown in Fig.~\ref{fig13}, from which we see that
$\sim99.8\,\,\%$ of the population is in the lower state during
der the
situation in which it is within the first allowed band 
the time inside the cavity, as predicted by the overlap. After
leaving the cavity, the atom regains its original internal and momentum state.

As mentioned above, scattering $N$ excited atoms against
an initially empty cavity, so that they are all reflected in the
lower state, creates a Fock state with exactly $N$ photons. Other
interesting situations may be realized as well with the same idea.
Assuming the cavity to be empty from the beginning and the atom in
some linear combination of upper and lower state and with a
momentum corresponding to the first gap, the state evolves as
\begin{equation}
\Big(a|\uparrow\rangle+b|\downarrow\rangle\Big)|q/2\rangle|0\rangle\,\Rightarrow\,\Big(a|-q/2,1\rangle+b|q/2,0\rangle\Big)|\downarrow\rangle.
\end{equation}
Since a lower state atom will traverse unaffected through the
cavity, while the upper state atom is reflected, this is realizing
an atomic 'Stern-Gerlach' type of measurement. If instead, the
field is initially in some linear combination of vacuum and the
one photon state and the atom in the lower state it follows that
\begin{equation}
|\downarrow,q/2\rangle\Big(a|0\rangle+b|1\rangle\Big)\,\Rightarrow\,\Big(a|\downarrow,q/2\rangle+b|\uparrow,-q/2\rangle\Big)|0\rangle.
\end{equation}
Thus a photon 'Stern-Gerlach' type of measurement is realized,
assuming that the field is restricted to the states $|0\rangle$
and $|1\rangle$. Obviously, the scheme may not be used only for
separating atomic or field stats, but it works also as a model for
creating entanglement between the involved degrees of freedom,
momentum, atomic and field states.


\section{Conclusion}
\label{s4} We have analyzed theoretically the problem of a
two-level atom propagating within a quantized standing wave
electromagnetic field, with an emphasis on the mass parameters in
atomic cavity QED systems. We have approached the problem with two
different strategies: the Floquet theory for the periodic Hamilton
operator, and the wave packet simulations. The first is a purely
mathematical  procedure, which offers numerical and approximate
analytic expressions for the mass parameters, while the second is
a more physical attack. The analytic results offer a hint of the
dependence of the masses on the physical parameters. Roughly
speaking, a small detuning $\Delta$, small photon momentum $q$,
and a large atom-field coupling $g_0$ yield large mass shifts.

The dynamics of an atomic wave packet inside (or entering) the
cavity can be understood in terms of the dispersion curve
$E^\nu(k)$, where $k$ is the quasi-momentum of a generalized plane
wave and $\nu$ is the band index. In the Taylor expansion around some
quasi-momentum $k_0$ each term defines a mass paramter; in this
paper we have considered the mass $m_1=\hbar k_0/v_{{\rm g}}$ related to
the group velocity $v_{{\rm g}}$ of Gaussian wave packets and the mass
$m_2$ associated with the spreading of the wave packet:
\begin{equation}
\frac{1}{m_1}=\left.\frac{1}{\hbar^2k_0}\frac{\partial E^\nu(k)}{\partial k}\right|_{k=k_0} \, \mathrm{and} \, \, \, \frac{1}{m_2}=\frac{1}{\hbar^2}\left.\frac{\partial^2E^\nu(k)}{\partial k^2}\right|_{k=k_0}.
\end{equation}
Note that $m_2$ is the same as the customary effective mass $m^*$
introduced to describe, for example, acceleration of electrons
within a crystal under an external force. These parameters vary
essentially over different scales of the coupling, as seen
in Fig.~\ref{fig6}. Even an arbitrary small coupling $g$ results in strong effects in the atomic dynamics. For $g\ge\frac{1}{2} $, both the group velocity and $1/m_2$ approaches zero and finally saturates towards this value, which in unscaled quantities can be written as
\begin{equation}
E_{\rm int} = \hbar\tilde{g} \le \frac{\hbar^2 \tilde{q}^2}{2m} = E_{\rm kin}.
\label{eq:upperbound}
\end{equation}
Hence, whenever the characteristic interaction energy exceeds the characteristic kinetic energy of
the lowest band, the band is essentially flat and the energy becomes nearly 
independent of the quasi-momentum. Using typical values $\lambda=1000$ nm. and $m=100$ a.m.u. we get a characteristic time scale $T_s\sim0.4$ $\mu s.$ giving us the unscaled coupling $\tilde{g}=g/T_s\approx10$ kHz for saturation. Note that dividing our dimensionless  detuning $\Delta$ and the coupling $g$ by $T_s$ gives real physical quantities.

The effective masses are only defined for dressed states with well defined momenta, and in many situations, the
system starts out in a bare state. It is, however, possible that
the bare states have a large overlap with some dressed state and
may be assigned an effective mass. This has been investigated in
great detail in the present paper and numerical simulation of
propagation for both Gaussian dressed and Gaussian bare states has
been analyzed and compared with both each other and with results
given by the Floquet theory. Both the group velocity and the
effective mass $m_2$ have been extracted numerically in the
various cases with a good agreement as compared with the values
obtained from the Floquet theory.

In most experiments today, the atom and field are prepared
separately and the atom is shot through the cavity with a
preselected velocity. The state of the complete system is then,
most likely, not in a dressed state, but rather a bare state. We
have not, in the text, discussed in detail how one may prepare
initial dressed states. A standard method is to use a slow
adiabatic change of some external controllable parameter. For the
JC model, bare and dressed states becomes identical in the two
limits $g_0\rightarrow0$ or $\Delta\rightarrow\pm\infty$. By
letting the atom traverse a Fabry-Perot cavity almost
perpendicular to the standing wave mode it is possible to have
very small initial atomic velocities $\hbar k_0/m$ along the
standing wave mode even for not so cold atoms by varying the angle
of injection; the atoms will have a high transverse velocity and a slow
velocity in the $x$-direction along the standing wave mode. Thus,
it is enough to quantize the atomic motion in just the
$x$-direction, while the $y$ and $z$-directions are treated
classically. In the case of a Fabry-Perot cavity, the transverse
Gaussian mode shapes can then be described by the time envelope $\exp(-t^2/\Delta t^2)$. If the turn-on of the transverse
Gaussian mode is slow enough, the atom will traverse the cavity in
a dressed state. Another feasible way to prepare the dressed
states is to use an external ion trap to confine the ion inside an
off-resonance cavity, and then slowly tune the cavity into
resonance and shut off the external trap.

In addition to investigating the dynamics of the atom in the
presence of the standing wave mode, we have suggested various
applications for state preparations and measurements. It has been
explained and shown numerically how an incident atom may be
reflected by the cavity while simultaneously leaving one photon
into it, which might be used for creation of Fock states. The
model can also be used for 'Stern-Gerlach' type of measurements,
separating the upper and lower atomic states spatially. Likewise,
it could also be used for separating the vacuum $|0\rangle$ from the
one photon state $|1\rangle$ and to create various entangled
states, like EPR-states.

The paper is purely theoretical, and we understand that experimental 
measurement of the effective masses will pose several challenges, 
depending on the approach chosen. First of all, the nodes of a 
free-space standing wave have a tendency to drift, which is avoided in 
a natural way by using an optical or a microwave cavity. Both regimes, 
however, bring new difficulties: A periodic field mode is needed in 
order to observe changes predicted in the propagation characteristics; 
present high-Q microwave cavities have, unfortunately, dimensions of 
the order of wavelength. Hence the optical wave number is not well 
defined and the Floquet approach adapted in this paper is no longer 
applicable~\cite{microwave}. In the optical regime, time scales of both 
the atomic spontaneous emission and the cavity decay are usually 
shorter than the interaction time, and both processes will strongly 
affect the atom-cavity system in a way that is not included in the 
present consideration. One way to go around these difficulties is to 
use extremely long lived transitions and configurations in which upper 
atomic levels are only slightly populated, cf. Figs.~\ref{fig6} 
and~\ref{fig7}; the cavity decay may be circumvented by using a driven 
cavity, hence rendering the light field effectively classical. Similar 
experiments are performed using classical laser fields made stable by a 
set of mirrors, see e.g.~\cite{bloch}. The same kind of setup could in 
principle be used for determination of the effective mass and one may then 
neglect decay of the field. The dynamics of the atom will then be the 
same when it interacts with a classical field as with a Fock state.

The Floquet approach presented for the description of atomic motion in 
terms of an effective mass parameters is, in principle, independent of 
weather the two-level atom interacts with a classical light field or 
with a photonic Fock state of a quantised field. The present results 
may, however, not be realistically achievable in today's cavity QED 
systems. Our treatment does set the stage for the concept of an 
electromagnetic effective mass and we may only hope that novel 
experimental conditions will, one day, make possible the observation of 
cavity-induced mass modification.


\section*{Acknowledgements}

JS acknowledges Quantum Complex Systems (QUACS)
Research Training Network (HPRN-CT-2002-00309)
for funding.


\begin{thebibliography}{99}

\bibitem{qed}\label{qed} {\it Cavity Quantum Electrodynamics, Advances in Atomic, Molecular and Optical Physics}, edited by P.R.~Berman (New york, Academic Press, 1994)

\bibitem{jc1} E.T.~Jaynes and F.W.~Cummings, Proc. \ IEEE, {\bf 51}, 89 (1963).

\bibitem{jc2} B.W.~Shore and P.L.~Knight, J. \ Mod. \ Opt. {\bf 40}, 1195 (1993).

\bibitem{qc1} A.~Rauschenbeutel, G.~Nogues, S.~Osnaghi, P.~Bertet, M.~Brune, J.M.~Raimond and S.~Haroche, \ Phys. \ Rev. \ Lett. {\bf 83}, 5166 (1999).

\bibitem{qc2} S.B.~Zheng and G.C.~Guo, \ Phys. \ Rev. \ Lett. {\bf 85}, 2392 (2000).

\bibitem{qc3} L.M.~Duan, A.~Kuzmich and H.J.~Kimble, \ Phys. \ Rev. A {\bf 67}, 032305 (2003).

\bibitem{qc4} J.~Larson and B.M.~Garraway, J. \ Mod. \ Opt. {\bf 51}, 1691 (2004).

\bibitem{fock} H.~Walther, J. \ Mod. \ Opt. {\bf 51}, 1859 (2004).

\bibitem{cat} M.~Brune, E.~Hagley, J.~Dreyer, X.~Maitre, A.~Maali, C.~Wunderlich, J.M.~Raimond and S.~Haroche, \ Phys. \ Rev. \ Lett. {\bf 77}, 4887 (1996).

\bibitem{mazer0} M.O.~Scully, G.M.~Meyer and H.~Walther, \ Phys. \ Rev. \ Lett. {\bf 76}, 4144 (1996).

\bibitem{mazer1} G.M.~Meyer, M.O.~Scully and H.~Walther, \ Phys. \ Rev. A {\bf 56}, 4142 (1997).

\bibitem{refl1} B.G.~Englert, J.~Schwinger, A.O.~Barut and M.O.~Scully, \ Europhys. \ Lett. {\bf 14}, 25 (1991).

\bibitem{refl2} M.~L\"offler, G.M~Meyer, M.~Schr\"oder, M.O.~Scully and H.~Walther, \ Phys. \ Rev. A {\bf 56}, 4153 (1997).

\bibitem{refl3} R.~Arun and G.S.~Agarwal, \ Phys. \ Rev. A {\bf 64}, 065802 (2001).

\bibitem{trap1} S.~Haroche, M.~Brune and J.M.~Raimond, \ Europhys. \ Lett. {\bf 14}, 19 (1991).

\bibitem{trap2} P.W.H.~Pinkse, T.~Fisher, P.~Maunz and G.~Rempe, \ Nature {\bf 404}, 365 (2000).

\bibitem{trap3} C.J.~Hood, T.W.~Lynn, A.C.~Doherty, A.S.~Parkins and H.J.~Kimble, \ Science {\bf 287}, 1447 (2000).

\bibitem{trap4} A.C.~Doherty, T.W.~Lynn, C.J.~Hood and H.J.~Kimble, \ Phys. \ Rev. A {\bf 63}, 013401 (2000).

\bibitem{trap4b} J.~Ye, D.W.~Vernooy and H.J.~Kimble, \ Phys. \ Rev. \ Lett. {\bf 83}, 4987 (1999).

\bibitem{trap5}C.~Sch\"on and J.I.~Cirac, \ Phya. \ Rev. A {\bf 67}, 043813 (2003).

\bibitem{trap6} P.~Horak, G.~Hechenblaikner, K.M.~Gheri, H.~Stecher and H.~Ritsch, \ Phys. \ Rev. \ Lett. {\bf 79}, 4974 (1997).

\bibitem{ext} D.W.~Vernooy and H.J.~Kimble, \ Phys. \ Rev. A {\bf 56}, 4287 (1997).

\bibitem{atcavtrap} A.B.~Mundt, A.~Kreuter, C.~Becher, D.~Leibfried, J.~Eschner, F.~Schmidt-Kaler and R.~Blatt, \ Phys. \ Rev. \ Lett. {\bf 89}, 103001 (2002).

\bibitem{dyn1} W.~Ren and H.J.~Carmichael, \ Phys. \ Rev. A {\bf 51}, 752 (1995).

\bibitem{dyn2}J.T.~Zhang, X.L~Feng, W.Q.~Zhang and Z.Z.~Xu, \ Chin. \ Phys. \ Lett. {\bf 19}, 670 (2002).

\bibitem{dyn3} Y.T.~Chough, S.H.~Youn, H.~Nha, S.W.~Kim and K.~An, \ Phys. \ Rev. A {\bf 65}, 023810 (2002).

\bibitem{dyn4} M.~Wilkens, E.~Schumacher and P.~Meystre, Phys. Rev. A {\bf 44}, 3130 (1991).

\bibitem{dyn5} I.~Cusumano, A.~Vaglica and G.~Vetri, \ Phys. \ Rev. A {\bf 66}, 043408 (2002).

\bibitem{dyn6} C.J.~Hood, M.S.~Chapman, T.W.~Lynn and H.J.~Kimble, \ Phys. \ Rev. \ Lett. {\bf 80}, 4157 (1998).

\bibitem{appdyn1} F.~Saif, F.~LeKien and M.S.~Zubairy, \ Phys. \ Rev. A {\bf 64}, 043812 (2001).

\bibitem{appdyn2} G.~Compagno, J.S.~Peng and F.~Persico, \ Phys. \ Rev. A {\bf 26}, 2065 (1982).

\bibitem{appdyn3} A.Z.~Muradyan and G.A.~Muradyan, \ J. \ Phys. B {\bf 35}, 3995 (2002).

\bibitem{appdyn4}A.~Vaglica, \ Phys. \ Rev. A {\bf 52}, 2319 (1995).

\bibitem{rm1} A.M.~Herkommer, V.M.~Akulin and W.P.~Schleich, \ Phys. \ Rev. \ Lett. {\bf 69}, 3298 (1992).

\bibitem{rm2} M.J.~Holland, D.F.~Walls and P.~Zoller, \ Phys. \ Rev. \ Lett. {\bf 67}, 1716 (1991).

\bibitem{rm3} P.~Storey, M.~Collett and D.~Walls, \ Phys. \ Rev. \ Lett. {\bf 68}, 472 (1992).

\bibitem{feyman}R. P.~Feyman, {\it Statistical Mechanics}, (Addison Wesley, 1990).

\bibitem{kittel} C.~Kittel, {\it Quantum Theory of Solids}, (New York, Wiley, 1987).

\bibitem{jonas2} J.~Larson and S.~Stenholm, J. \ Mod. \ Opt. {\bf 50}, 2705 (2003).

\bibitem{schlicher} R.R.~Schlicher. Opt. \ Commun. {\bf 70}, 97 (1988).

\bibitem{stigadd1} A.C.~Doherty, A.S.~Parkins, S.M.~Tan and D.F.~Walls, \ Phys. \ Rev. A {\bf 57}, 4804 (1998).

\bibitem{stigadd2} T.~Sleator and M.~Wilkens, \ Phys. \ Rev. A {\bf 48}, 3286 (1993).

\bibitem{math} {\it Handbook of Mathematical Functions}, Natl. Bur. Stand. Appl. Math. Ser. No. 55, edited by M.~Abramowitz and I.A.~Stegun (U.S. GPO, Washington, DC, 1972).

\bibitem{kolovsky} A.~R.~Kolovsky, J.~Opt.~B {\bf 4}, 218 (2002).

\bibitem{stigadd3}E.~Arimondo, A.~Bambini and S.~Stenholm, Phys. Rev. A {\bf 24}, 898 (1981).

\bibitem{split} M.D.~Fleit, J.A.~Fleck, and A.~Steiger, J. Comp. Phys. {\bf 81}, 412 (1982).

\bibitem{doplaser} M.~Brune, E.~Hagley, J.~Dreyer, X.~Maitre, A.~Maali, C.~Wunderlich, J.M.~Raimond and S.~Haroche, \ Phys. \ Rev. \ Lett. {\bf 77}, 4887 (1996).

\bibitem{microwave} D.~Meschede, H.~Walther, and G.~M\"uller, \ Phys. \ Rev. \ Lett. {\bf 54}, 551 (1985).

\bibitem{bloch} Q.~Niu, X-G.~Zhao, G. A.~Georgakis, and M. G.~Raizen, \ Phys. \ Rev. \ Lett. {\bf 76}, 4504 (1996).


 \end{thebibliography}
\end{document}